\documentclass[manuscript]{emulateapj}
\usepackage{graphics}
\usepackage{graphicx}
\usepackage{rotating}
\usepackage{mathtools}
\usepackage{footmisc}
\begin{document}

\newcommand{\kms}{\mbox{km~s$^{-1}$}}
\newcommand{\s}{\mbox{$''$}}
\newcommand{\mloss}{\mbox{$\dot{M}$}}
\newcommand{\mdot}{\mbox{$\dot{M}$}}
\newcommand{\my}{\mbox{$M_{\odot}$~yr$^{-1}$}}
\newcommand{\ls}{\mbox{$L_{\odot}$}}
\newcommand{\um}{\mbox{$\mu$m}}
\newcommand{\ujy}{\mbox{$\mu$Jy}}
\newcommand{\ms}{\mbox{$M_{\odot}$}}

\newcommand{\vexp}{\mbox{$V_{\rm exp}$}}
\newcommand{\vsys}{\mbox{$V_{\rm sys}$}}
\newcommand{\vlsr}{\mbox{$V_{\rm LSR}$}}
\newcommand{\tex}{\mbox{$T_{\rm ex}$}}
\newcommand{\teff}{\mbox{$T_{\rm eff}$}}
\newcommand{\tmb}{\mbox{$T_{\rm mb}$}}
\newcommand{\trot}{\mbox{$T_{\rm rot}$}}
\newcommand{\tkin}{\mbox{$T_{\rm kin}$}}
\newcommand{\dens}{\mbox{$n_{\rm H_2}$}}
\newcommand{\bri}{\mbox{erg\,s$^{-1}$\,cm$^{-2}$\,\AA$^{-1}$\,arcsec$^{-2}$}}
\newcommand{\brib}{\mbox{erg\,s$^{-1}$\,cm$^{-2}$\,arcsec$^{-2}$}}
\newcommand{\flux}{\mbox{erg\,s$^{-1}$\,cm$^{-2}$\,\AA$^{-1}$}}
\newcommand{\ha}{\mbox{H$\alpha$}}

\title{High-Speed Bullet Ejections during the AGB to Planetary Nebula Transition: HST Observations of the Carbon Star, V Hydrae}
\author{R. Sahai\altaffilmark{1}, S. Scibelli\altaffilmark{1,2}, \& M. R. Morris\altaffilmark{3}}

\altaffiltext{1}{Jet Propulsion Laboratory, MS\,183-900, California Institute of Technology, Pasadena, CA 91109, USA}
\altaffiltext{2}{Stony Brook University, Stony Brook, NY 11794-3800, USA}
\altaffiltext{3}{Department of Physics and Astronomy, UCLA, Los Angeles, CA 90095-1547}

\email{raghvendra.sahai@jpl.nasa.gov}
\begin{abstract}
The well-studied carbon star, V\,Hya, showing evidence for high-speed, collimated outflows and dense equatorial structures, is 
a key object in the study of the poorly understood transition of AGB stars into aspherical planetary nebulae. 
Using the STIS instrument onboard HST, we have obtained high spatial-resolution long-slit optical spectra of V\,Hya that 
show high-velocity emission in [SII] and [FeII] lines. Our dataset, spanning three epochs spaced apart by a year during 
each of two periods (in 2002-2004 and 2011-2013), shows that V\,Hya ejects high-speed ($\sim200-250$\,\kms) bullets  
once every $\sim$8.5\,yr. The ejection axis flip-flops around a roughly eastern direction, both in and perpendicular to the sky-plane, 
and the radial velocities of the ejecta also vary in concert between 
low and high values. We propose a model in which the bullet ejection is associated with the periastron passage of a binary 
companion in an eccentric orbit around V\,Hya with an orbital period of $\sim$8.5\,yr. The flip-flop phenomenon is likely the result of 
collimated ejection from an accretion disk (produced by gravitational capture of material from the primary) that is warped and precessing, 
and/or that has a magnetic field that is misaligned with that of the companion or the primary star. We show   
how a previously observed 17-year period in V\,Hya's light-cycle can also be explained in our model. Additionally, we 
describe how the model proposed here can be extended to account for multipolar nebulae.
\end{abstract}


\section{Introduction}\label{intro}
Most knwon planetary nebulae (PNe) are not round (with more than half being bipolar or multipolar), and there are no round pre-planetary nebulae (PPNe), 
as revealed by imaging surveys of young PNe with HST 
(e.g., Sahai \& Trauger 1998 [hereafter ST98], Sahai et al. 2011, Sahai et al. 2007), whereas the dense mass loss that occurs during the 
progenitor AGB phase is mostly spherical. The widespread presence of point-symmetry in PN morphologies led to ST98's proposal that high-speed, collimated (jet-like) outflows that can change their orientation and are initiated during the very-late AGB phase, sculpt the spherical AGB mass-loss envelopes from the inside out, producing the observed variety of aspherical shapes.
These jets are perhaps driven by an accretion disk around a
binary companion (Morris 1987), but direct evidence for such jets and disks in AGB stars
is rare. 

The carbon star, V\,Hya, is one object where we have evidence for both high-speed, collimated outflows and a ``disk". 
Early evidence for the presence of very high-speed outflows came from observations of CO
4.6\micron~vibration-rotation lines in V\,Hya (Sahai \& Wannier 1988). A follow-up, multi-epoch 
study revealed the physical properties of multiple high-velocity outflow components 
with $V_{exp} \approx 70-120$ \kms, and found significant time variability in these, even over time-scales as short as one day  
(Sahai, Sugerman \& Hinkle 2009). Lloyd Evans (1991) obtained optical spectra in the blue-violet range and 
found emission lines indicating outflow velocities up to about $160$\,\kms. High signal-to-noise 
millimeter-wave CO J=2-1 and 3-2 line profiles showed wide wings, implying that the
high-speed ($200$\,\kms) outflow was fairly massive (Knapp
et al. 1997). Mm-wave interferometry of the CO lines by Hirano et al. (2004) with $\sim3{''}-4{''}$ resolution showed that 
the high-velocity outflow is bipolar, blobby and highly collimated (see also Kahane et al. 1996). 

Using the STIS instrument onboard HST, Sahai et al. (2003: SMKYB03) discovered a newly launched,
high-speed blob in V\,Hya via emission in the [SII]$\lambda\lambda\,4069.7,4077.5$ doublet. The blob had a 
projected radial velocity\footnote{All 
radial velocities are given in the heliocentric reference frame; for comparison with millimeter-wave studies where velocities 
are usually given in the LSR reference frame, $V_{lsr}=V_{hel}-11$\,\kms} of about -250\,\kms, and was offset by $0\farcs25$ 
from the central star (100\,AU at V\,Hya's estimated distance, D=400\,pc: Kahane et al. 1996, Knapp et al. 1999),
and had a measurable proper motion of $0\farcs07$ yr$^{-1}$. SMKYB03 also found a hot, central disk-like structure of diameter $0\farcs6$ 
expanding at a speed of $10-15$\,\kms. 

The disk-like structures seen in H$\alpha$ and CO emission are large and 
expanding, and therefore not the result of
accretion. SMKYB03 conjecture that these may have resulted from a recent phase of equatorially-enhanced mass-loss
in V\,Hya, possibly as a result of V Hya being 
a binary in a common-envelope configuration, as proposed by Barnbaum, Morris, \& Kahane (1995).

The accretion disk itself is likely to be seen in the UV and in X-rays. 
Sahai et al. (2008) found that V\,Hya belongs to the newly-discovered 
``fuvAGB" class of AGB stars that show variable FUV emission, likely produced by actively-accreting companions. 
Observations of a small sample of fuvAGB stars show 
variable X-ray emission from very hot gas, that is likely related to hot, active accretion 
disks around the companion stars, although V\,Hya itself was not detected in this survey (Sahai et al.\,2015). The non-detection 
of X-ray emission from V\,Hya  
may be related to the fact that it depends on the possibly variable accretion rate. The X-ray observations were done long after ($\sim9$\,yr) 
the UV observations, and given the variability, we do not have any constraint on if, and how, the UV and X-ray emission in V\,Hya, may be 
coupled.


In this paper, we present new STIS observations that span more than a decade, and characterize for the first time 
the extended history of ejection of high-speed clumps (bullets) from this object. The paper is organized as follows. In \S\,\ref{obsreduce}, we
summarize the observational setup and data-reduction, and in \S\,\ref{obsresult}, we provide our main observational results. In the discussion 
section (\S\,\ref{discuss}), we first infer the detailed history and characteristics of the bullet ejections in V\,Hya 
(\S\,\ref{history}), followed by an analysis of the 
physical conditions in the bullets (\S\,\ref{blobphys}). We end the discussion with a binary model that can explain our main 
observational results as well as previous observations (\S\,\ref{binmodel}). Concluding remarks are provided in \S\,\ref{conclude}.

\section{Observations and Data Reduction}\label{obsreduce}
We obtained long-slit spectroscopy of V Hya with the Space Telescope Imaging Spectrograph (STIS) onboard the Hubble Space 
Telescope (HST) during 6 epochs in two sets of 3 each. The first set, epochs 1--3 (collectively referred to as Period 1) covered the 
3-year period 
2002, 2003, 2004 (programs GO 9100, 9632, and 9800, respectively). The second set, epochs 4--6 
 (collectively referred to as Period 2) covered the 3-year period 2011, 2012, 2013 (programs GO 12227, 12664, 13053, respectively). 
Some of the data from GO 9100 and 9632 were reported by SMKYB03. 

\begin{figure}
\includegraphics[width = 0.9\linewidth]{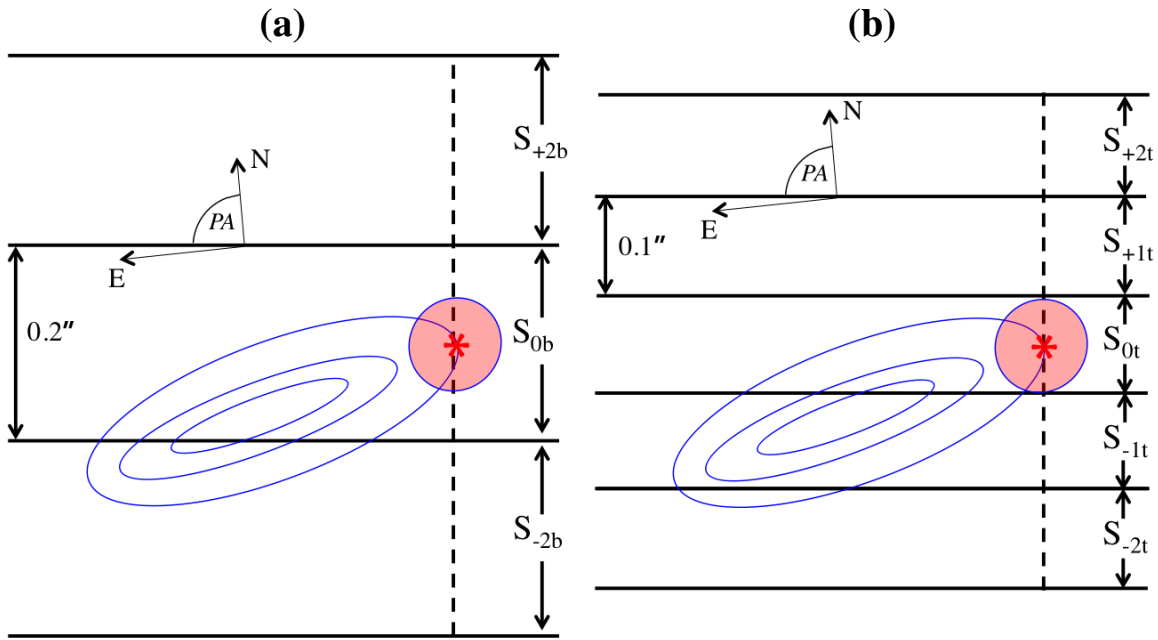}
\caption{Schematic representation of the slit mosaics used for different epochs. (a) A mosaic of broad slits (i.e., with width=$0\farcs2$), 
used for epochs 1, 4, 5, \& 6. (b) A mosaic of thin slits (i.e., with width=$0\farcs1$), used for epochs 2 \& 3 
(in epoch 3, slit $S_{-2t}$ was not used). The slits 
are oriented within a few degrees of the E--W direction (i.e., $PA=90\arcdeg$) (the misalignment has been exaggerated in this figure for clarity). 
The red circle represents the star and the blue elliptical 
shape represents bullet 1, seen as a detached blob in Period 1 (not to scale). 
}
\label{slitmosaic}
\end{figure}

The pipeline-calibrated STIS [SII] spectra were retrieved from the MAST HST archive. 
These spectra have spaxels of size $0.276\AA\times0.051{''}$, and cover the wavelength range $4050-4359$\,\AA.
A log of the observations is given in Table\,\ref{obslog}. During each epoch, we used a mosaic of 3, 4, or 5 slits, 
roughly aligned east-west, covering locations on the central source, and offset north and south from it. For each slit, the star was dithered 
along the slit over 3 locations (except for GO 9100, when this number was 2)
In the first epoch, January 2002, the slit width was
$0\farcs2$ and the slits were spaced $0\farcs2$ apart. Data for the next two epochs were taken in December 2002 
and January 2004, with slit widths of $0\farcs1$, spaced $0\farcs1$ apart. During July 2011, 2012 and 2013, a slit width of $0\farcs2$ was used 
and each slit was spaced $0\farcs2$ apart. The on-center slit, with a width of $0\farcs2$ ($0\farcs1$), is labeled
$S_{0b}$ ($S_{0t}$): the symbols ``b" or ``t" signify whether a 
broad ($0\farcs2$) or thin ($0\farcs1$) slit was used. The off-center slits are named similarly, with $0$ replaced by 
either $-2,-1,+1,$ or $+2$ as appropriate: the value of the numeral is the offset of the slit in arcsec multiplied by 10, and the 
sign of the numeral indicates whether the slit is north (+ve) or south (-ve) of center. A schematic representation of the broad and 
narrow 
slit mosaics is show in Fig.\,\ref{slitmosaic}.

\subsection{Data Reduction}
The data were reduced using IRAF. Bad pixel mask files were generated to remove hot pixels from the individual pipeline-calibrated spectra. 
Each of the cleaned spectra were subsampled by a factor 2 along the spatial direction. These individual dithered spectra were registered to a common 
spatial reference frame and averaged.  


\section{Observational Results}\label{obsresult}
\begin{figure}
\includegraphics[width = 1.0\linewidth]{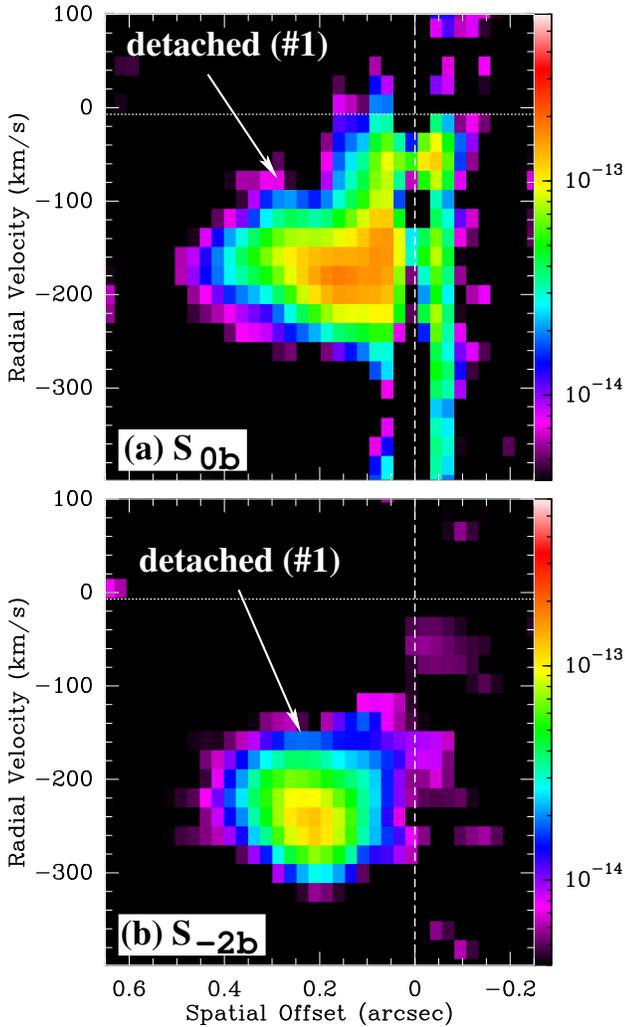}
\caption{STIS longslit spectra of V\,Hya showing the [SII]$\lambda4069.7$ line emission in a $0\farcs2$ slit from Epoch 1, 2002-01-28 (GO 
9100), oriented at $PA=-90\arcdeg$, and centered at offsets of (a) 0 (slit $S_{0b}$), and (b) $-0\farcs2$ (slit $S_{-2b}$) relative to the 
central star, 
along an axis oriented orthogonal to the slit. The central star 
is located at spatial offset 0 (dashed vertical line), and positive spatial offsets lie to its east. V\,Hya's systemic radial 
velocity, 
is $V_{hel}=-7$\,\kms~(horizontal dotted line). A background continuum has been subtracted. 
The scale color-bar shows intensities in units of erg\,s$^{-1}$\,cm$^{-2}$\,{\AA}$^{-1}$\,arcsec$^{-2}$. The dominant emission blob 
is labelled with its type (one-source, detached or distant), together with the number of the associated bullet.
}
\label{9100obs}
\end{figure}

\begin{figure}
\includegraphics[width = 1.0\linewidth]{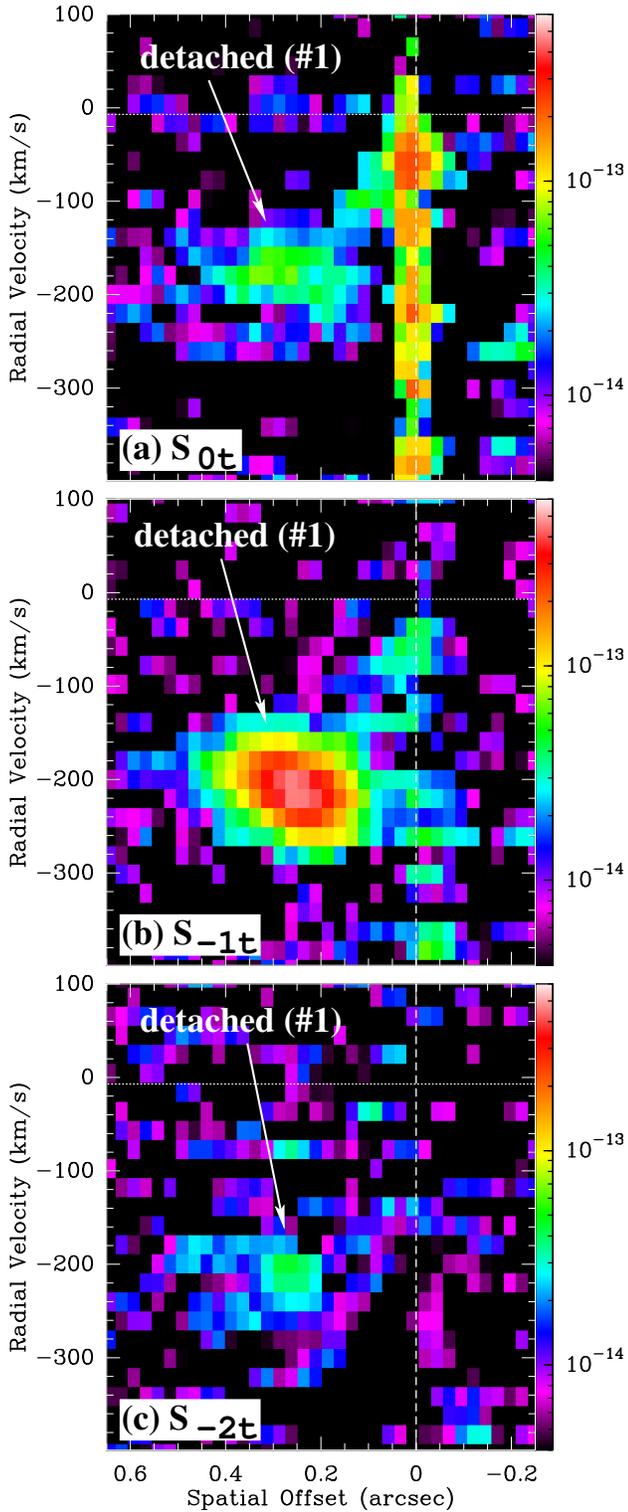}
\caption{As in Fig.\,\ref{9100obs}, but for Epoch 2, 2002-12-29 (GO 9632), with a $0.1{''}$ slit oriented at $PA=-92\arcdeg$, and centered 
at offsets of 
(a) 0 (slit $S_{0t}$), (b) $-0.1{''}$ (slit $S_{-1t}$), and (c) $-0\farcs2$ (slit $S_{-2t}$) relative to the central star.
}
\label{9632obs}
\end{figure}

\begin{figure}[htb]
\includegraphics[width = 1.0\linewidth]{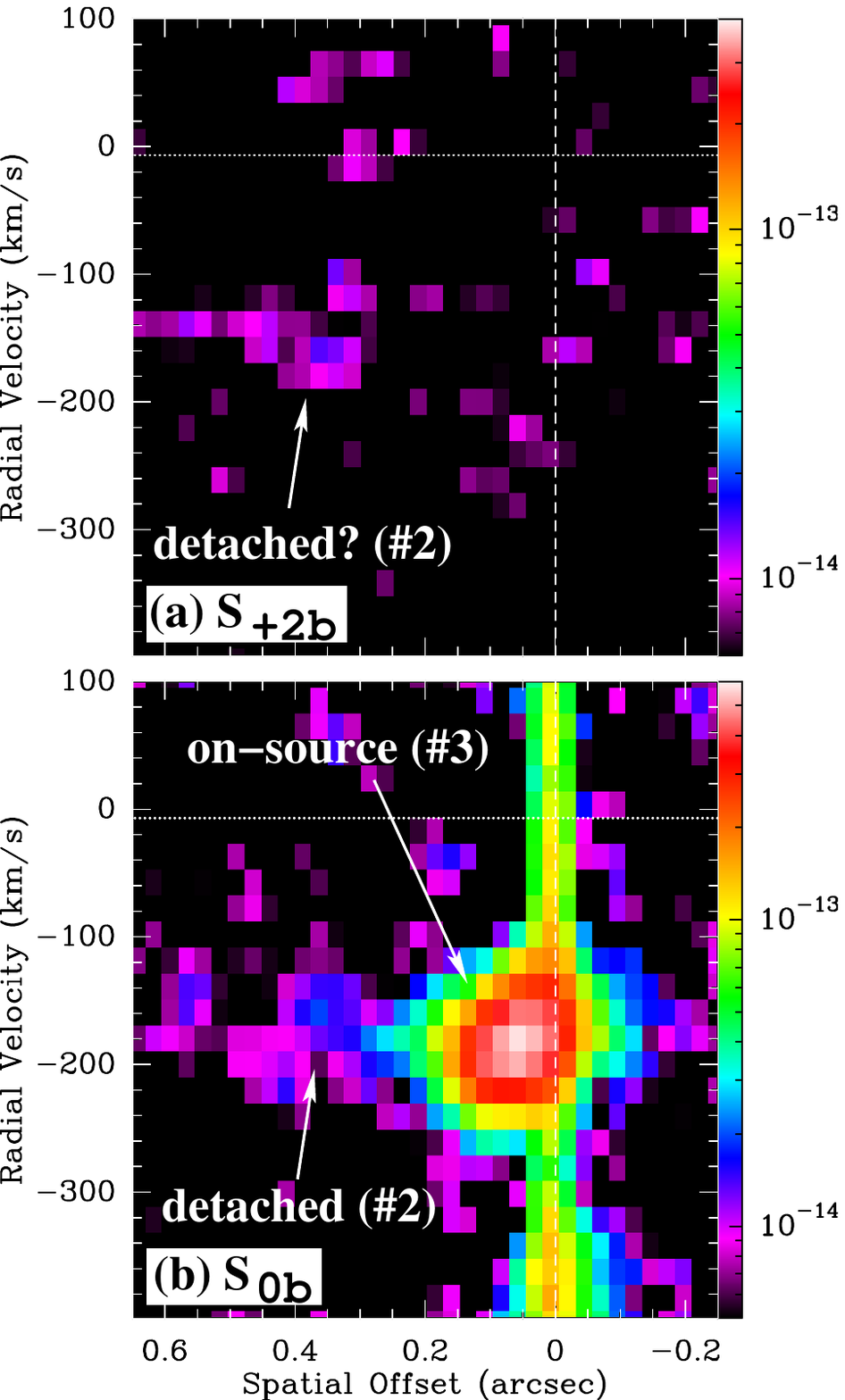}
\caption{As in Fig.\,\ref{9100obs}, but for Epoch 6, 2013-07-17 (GO 13053), with slit oriented at $PA=85\arcdeg$, and centered at offsets 
of 
(a) $0\farcs2$ (slit $S_{+2b}$), and (b) 0 (slit $S_{0b}$), 
relative to the central star.
}
\label{go13053-bg}
\end{figure}

\begin{figure}[htb]
\includegraphics[width = 0.47\linewidth]{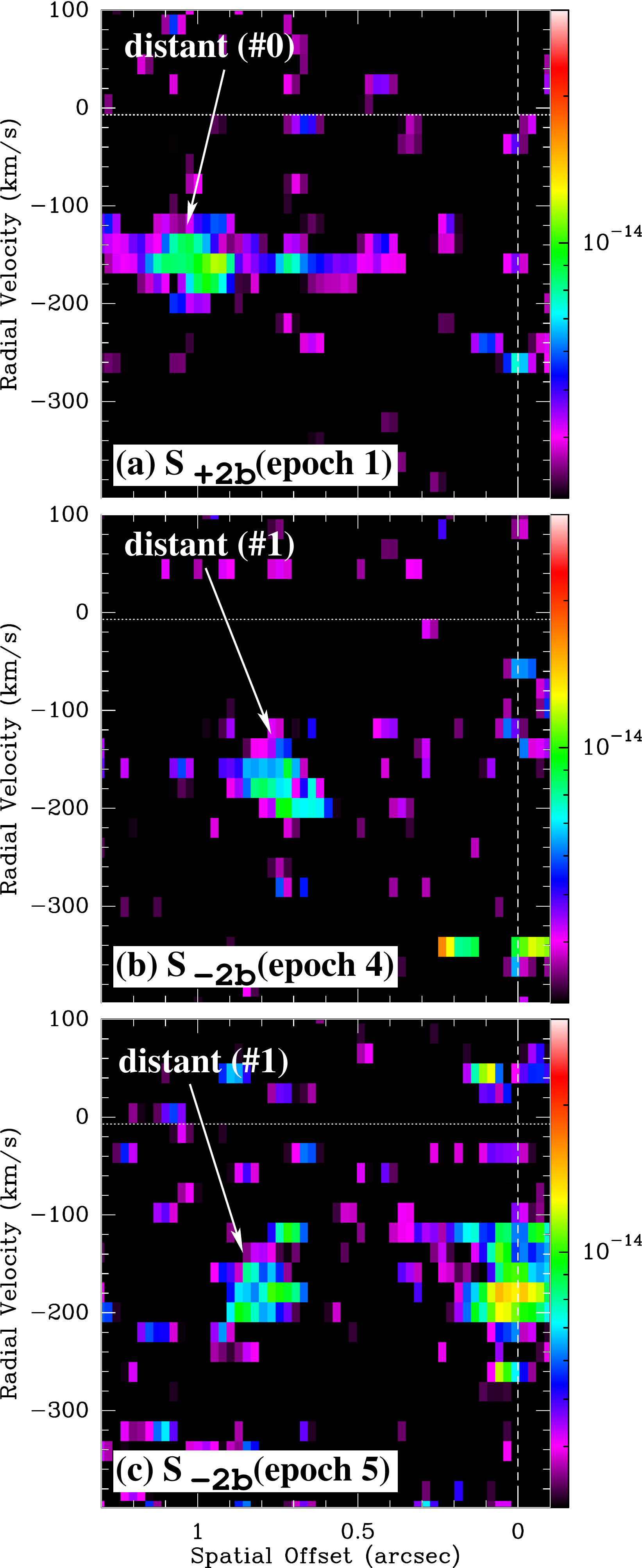}
\caption{As in Fig.\,\ref{9100obs}, but showing the presence of a distant blob. Panel (a) shows Epoch 1, with slit centered at 
an offset of $0\farcs2$ (slit $S_{+2b}$) relative to the central star, whereas panels (b) and (c) shows epochs 4 and 5, with slit centered 
at 
an offset of $-0\farcs2$ (slit $S_{-2b}$) relative to the central star.
}
\label{per1-per2-dist}
\end{figure}

During all epochs, we generally observe prominent emission in the [SII]$\lambda4069.7$ line (the stronger component of the doublet) 
from
one or both of two regions (Figs.\,\ref{9100obs}--\ref{go13053-bg}). The one that is seen most frequently 
is separated by about $0\farcs15$ to $0\farcs3$ from the continuum, towards the east (hereafter ``detached" 
emission/blob). A second one is centered 
on or very close to the continuum (hereafter ``on-source" emission/blob) but is not seen in all the epochs. 
A 3rd region of faint emission is sometimes seen at larger separations of about $0\farcs8$ to $1\farcs0$ 
(hereafter ``distant" emission/blob) (Fig.\,\ref{per1-per2-dist}). 

Important observational characteristics (spatial offset from center, radial velocity, linewidth and intensity) 
of the on-source and detached blobs as seen in [SII]$\lambda4069.7$ emission are given in Table\,\ref{obssii} 
as a function of epoch and slit. We also report the continuum intensity, as measured in a small 
line-emission-free wavelength window between the lines of the [SII] doublet. For the distant blob, which is much fainter and 
only detected during 3 epochs, the spatial offset, radial velocity, 
and intensity are provided in Table\,\ref{obsdist}. These same quantities  
are presented for the emission in the strongest two [FeII] lines, at 4245.16$\AA$ and 4288.6$\AA$ respectively, from on-source and 
detached blobs, in Tables\,\ref{obsfe1} and \ref{obsfe2}. The [FeII] data are given for the detached blob only for a 
single offset slit because the emission in the other slits is too weak (or undetected). For the same reason, 
we give the properties of the [FeII] line emission only for the on-source blob in the central slit in Period 2.

The spatial offset was determined with a Gaussian fit to the emission blob using the IRAF ``n2gaussfit" task. 
In order to find the radial velocity of peak emission, $V_P$, the linewidth (fwhm) and the fluxes, we (i) used the IRAF task ``pvector" to average the 
data spatially over three rows ($0\farcs076$)
covering the peak blob emission in the STIS images, and (ii) the IRAF task ``splot" to do a Gaussian profile-fit. For deriving the 
properties of the distant blobs (Table\,\ref{obsdist}), we averaged over nine rows ($0\farcs228$). 
For the [FeII] data in Table\,\ref{obsfe1} and Table \,\ref{obsfe2}, we averaged over five rows ($0\farcs127$), as the line intensities are lower than [SII]. The 
uncertainties in the $V_P$ values are relatively small (few\,\kms) for our high S/N detections of the [SII]$\lambda4069.7$ line, and somewhat higher 
($\sim10$\,\kms) for the [FeII] lines, where the S/N is lower.

In epoch 1, we see a detached blob at an offset of $0\farcs185$ ($0\farcs16$) in the $S_{-2b}$ ($S_{0b}$) 
slit (Fig.\,\ref{9100obs}), and a distant blob at an offset $1{''}$ in the $S_{+2b}$ slit (Fig.\,\ref{per1-per2-dist}a). The latter, which is rather faint, is not seen in epochs 2 and 3. All of these emission blobs are located east of the central 
source. During Period 1, the detached blob's proper motion is seen clearly 
in the slits offset to the south of the star. Since the slit width is $0\farcs2$ in 
epoch 1, and $0\farcs1$ in epochs 2 and 3, a direct determination of the proper motion of this blob 
from epoch 1 to epoch 2 cannot be made from a simple comparison of the blob locations in these two 
epochs. 

For epochs 1 and 2, the central slit also shows part of the detached blob that is closer to the center 
(Figs.\,\ref{9100obs}a,\,\ref{9632obs}a). In epoch 2, this detached-blob 
emission seen through the central slit is 
much weaker, and its centroid is displaced significantly further east from the continuum, compared to epoch 
1 -- a combined result of the proper motion of the blob and the reduction in slit-width. 

In epoch 3, the central slit shows relatively faint emission from the detached blob at an  
offset of roughly $0\farcs3$ to $0\farcs4$ (Fig.\,\ref{9800obs}a). 
In addition, a relatively bright on-source blob can also be seen in this slit, located just slightly east of 
the continuum (offset $0\farcs06$); this emission signifies a newly-ejected blob.

During epochs 4 and 5, slits $S_{+2b}$ clearly show emission from a detached 
blob at offsets of $0\farcs24$ and $0\farcs31$ (Figs.\,\ref{go12227-bg}a,\,\ref{go12664-bg}a). 
This emission blob is only marginally visible in epoch 6 in slit $S_{+2b}$ as it is much fainter, 
located roughly at an offset of $0\farcs36$ (Fig.\,\ref{go13053-bg}a). It is significantly brighter and 
clearly visible in slit $S_{0b}$ (Fig.\,\ref{go13053-bg}b), at an offset of about $0\farcs36$; it is likely present 
in the $S_{0b}$ slits in epochs 4 and 5 as well, but not sufficently well-separated from the bright on-source blob emission.
Thus, as during epochs 1--3, we find clear proper motion 
of the detached blob during epochs 5--6. A bright on-source emission blob is seen in 
the $S_{0b}$ slit for all 3 epochs (Figs.\,\ref{go12227-bg}b,\,\ref{go12664-bg}b,\,\ref{go13053-bg}b); this blob 
peaks in brightness during epoch 5. Although, as in Period 1, emission from the detached blob is also seen in 
the $S_{0b}$ slits, it is not well separated from the outer parts of the on-source blob.

A faint distant blob can be seen at an offset of about $0\farcs75$ ($0\farcs8$) in the $S_{-2b}$ slit during 
epoch 4 (epoch 5) (Fig.\,\ref{per1-per2-dist}b,c). 
It is not seen in epoch 6. The radial velocity of this blob, averaging the data from epochs 4 and 5, and over a spatial region of  
0\farcs178, is $-172$\,\kms. In comparison, the radial velocity of the distant blob in epoch 3, is $-161$\,\kms.





\section{Discussion}\label{discuss}
\subsection{Extended History of High-Velocity Blob Ejection}\label{history}
The presence of on-source, detached and distant emission-line blobs, as described in the previous section, indicates that 
V\,Hya is ejecting compact clumps of high-speed material, or bullets\footnote{henceforth, we use the word ``bullet" to represent the physical 
object that is ejected, and ``blob" to refer to its observational manifestation in our STIS data}. 
In this section, we use our data to determine the detailed history of these 
ejections over a time interval of about 25\,yr.

\begin{figure}
\includegraphics[width = 1.0\linewidth]{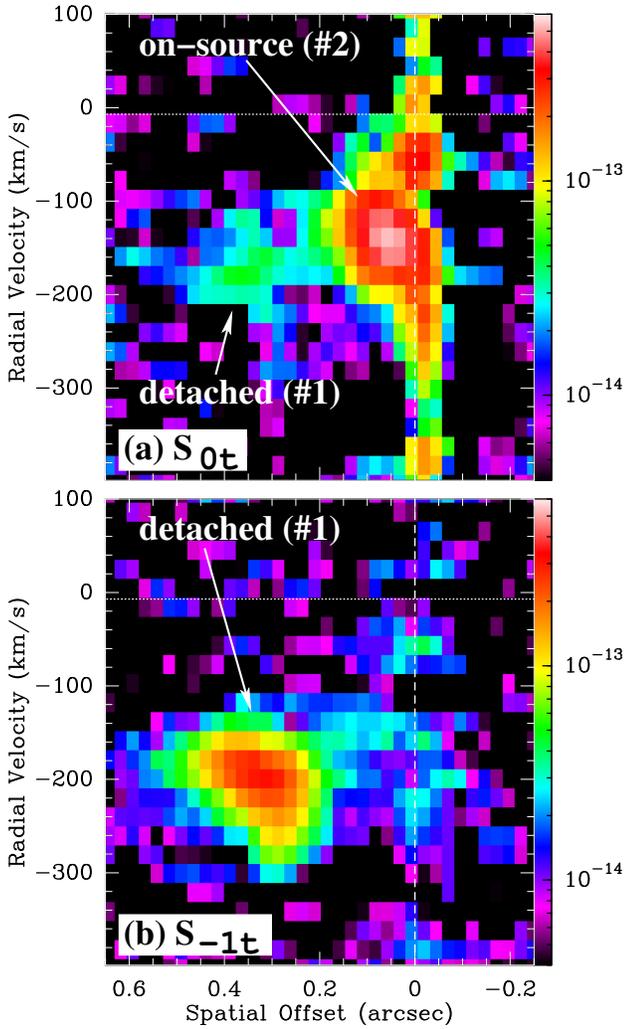}
\caption{As in Fig.\,\ref{9100obs}, but for Epoch 3, 2004-01-12 (GO 9800), with a $0.1{''}$-broad slit oriented at $PA=-93\arcdeg$, 
and centered at offsets of (a) 0 (slit $S_{0t}$), and (b) $-0.1{''}$ (slit $S_{-1t}$) relative to the central star. 
}
\label{9800obs}
\end{figure}

\begin{figure}[htb]
\includegraphics[width = 1.0\linewidth]{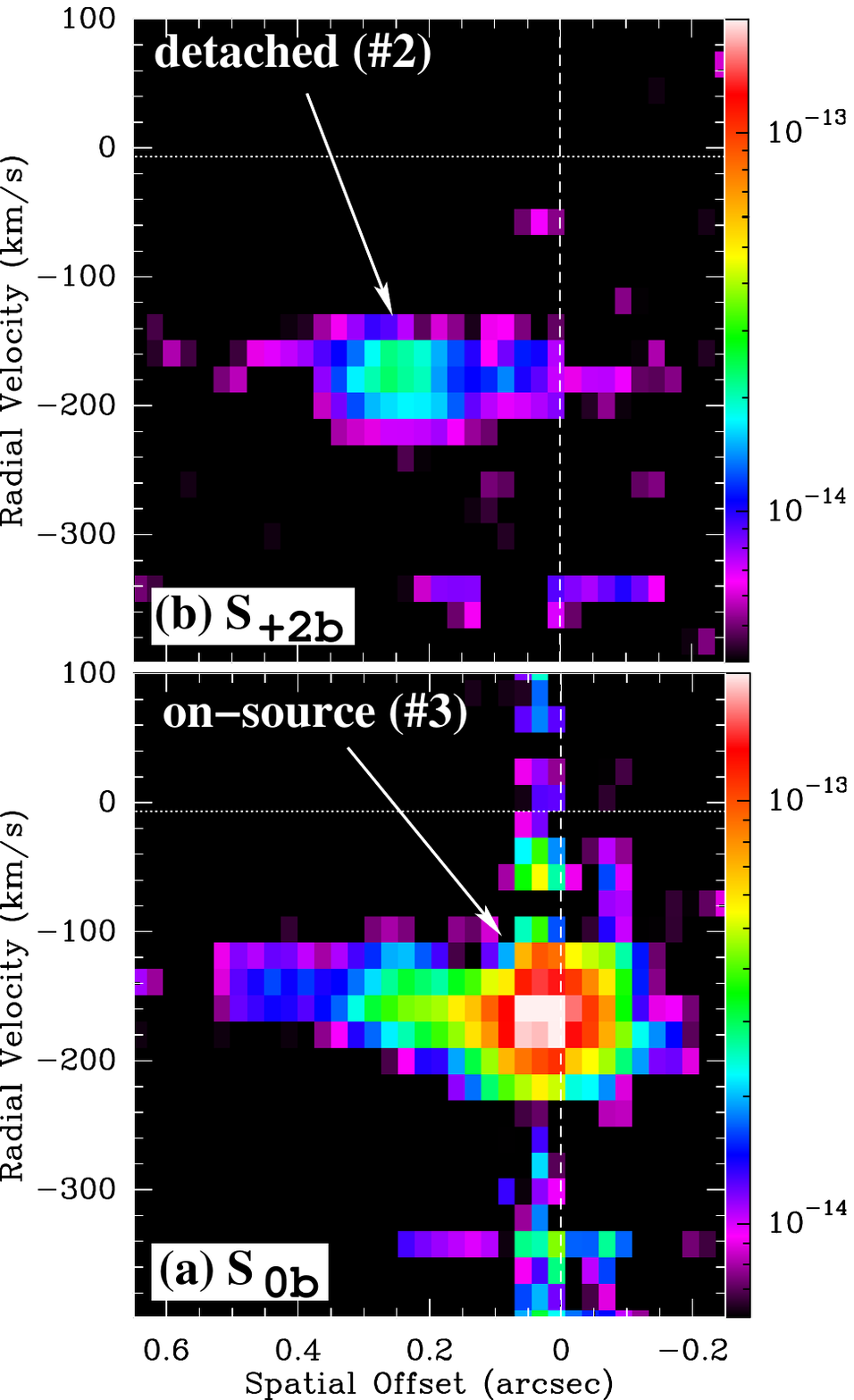}
\caption{As in Fig.\,\ref{9100obs}, but for Epoch 4, 2011-07-07 (GO 12227), with slit oriented at $PA=84\arcdeg$, and centered at offsets 
of 
(a) $0\farcs2$ (slit $S_{+2b}$), and (b) 0 (slit $S_{0b}$), relative to the central star.
}
\label{go12227-bg}
\end{figure}

\begin{figure}[htb]
\includegraphics[width = 1.0\linewidth]{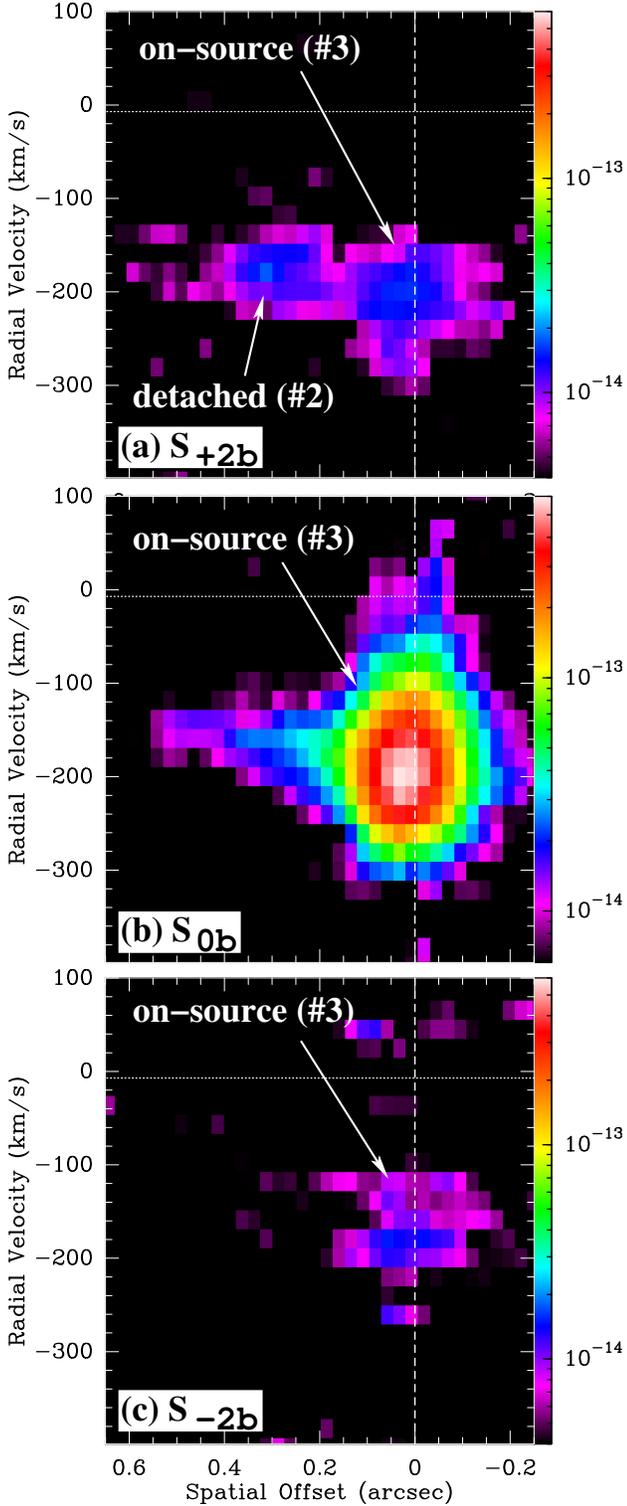}
\caption{As in Fig.\,\ref{9100obs}, but for Epoch 5, 2012-07-14 (GO 12664), with slit oriented at $PA=84\arcdeg$, and centered at offsets 
of 
(a) $0\farcs2$ (slit $S_{+2b}$), (b) 0 (slit $S_{0b}$), and (c) $-0\farcs2$ (slit $S_{-2b}$), 
relative to the central star.
}
\label{go12664-bg}
\end{figure}

During Period 1, an on-source blob is seen in epoch 3 (Fig.\,\ref{9800obs}a), and in Period 2, an on-source blob is seen in 
epochs 4--6 (Figs.\,\ref{go12227-bg}b,\,\ref{go12664-bg}b,\,\ref{go13053-bg}b): 
these represent newly-ejected bullets. 
The spatial offset of the on-source blob in epoch 3 is larger than that in epochs 4 and 5, and less than 
that in epoch 6. It is thus likely that the bullet-ejection in Period 1 occurred sometime before epoch 3, 
but the bullet became visible only in epoch 3 (see \S\,\ref{binmodel}). Thus, assuming a regular period, $P$, for bullet ejections, 
we get $P=P_0/N$, where $P_0\sim7.5-9.5$\,yr and N is some integer.

If $N=2$, then $P\sim4\,yr$, which would imply that the bullet ejection prior to the one seen in epoch 3 occured during  
$1998-2000$. If so, we can estimate where this bullet would be located in Period 1 by applying to it 
the ``early" history of blob movement as seen in Period 2.  
During Period 2, 
we find that the newly-ejected blob is located at an offset\footnote{we use the centroid of the peak 
[SII]$\lambda4069.7$ emission to measure the blob location} of $0\farcs033$ from the center in 
epochs 4 and 5 (Figs.\,\ref{go12227-bg}b,\,\ref{go12664-bg}b), and at an offset of $0\farcs071$ in epoch 6 
(Fig,\,\ref{go13053-bg}b). Thus, there is no significant proper motion during epochs 4--5, 
and then, from epoch 5 to 6, we see a proper 
motion away from the center of about $0\farcs038$. Hence emission from a new bullet ejected sometime during $1998-2000$  
would be seen as a bright, detached blob centered at an offset of about $0\farcs07$ sometime during Period 1, 
in the velocity range $-180$ to $-240$\,\kms. 
But since no such blob emission is seen in any of these epochs 
(see Figs.\,\ref{9100obs}b,\,\ref{9632obs}b\,\ref{9800obs}b), we conclude that N$\ne$2. 

Of course, N cannot be greater than two, as that would imply the presence of two or more blobs between the center and the 
detached blob in Period 1, contrary to what we observe.
Hence, we conclude that N=1, $P=P_0$, and the ejection prior to the one in Period 1 occurred sometime 
during $1993.5-1995.5$: the bullet ejected then (bullet 1) is seen as 
the detached blob in epochs 1, 2 and 3 (Figs.\,\ref{9100obs},\,\ref{9632obs},\,\ref{9800obs}), and the distant 
blob in epochs 4 and 5 (Fig.\,\ref{per1-per2-dist}b,c).

\begin{figure}[htb]
\resizebox{1.0\textwidth}{!}{\includegraphics{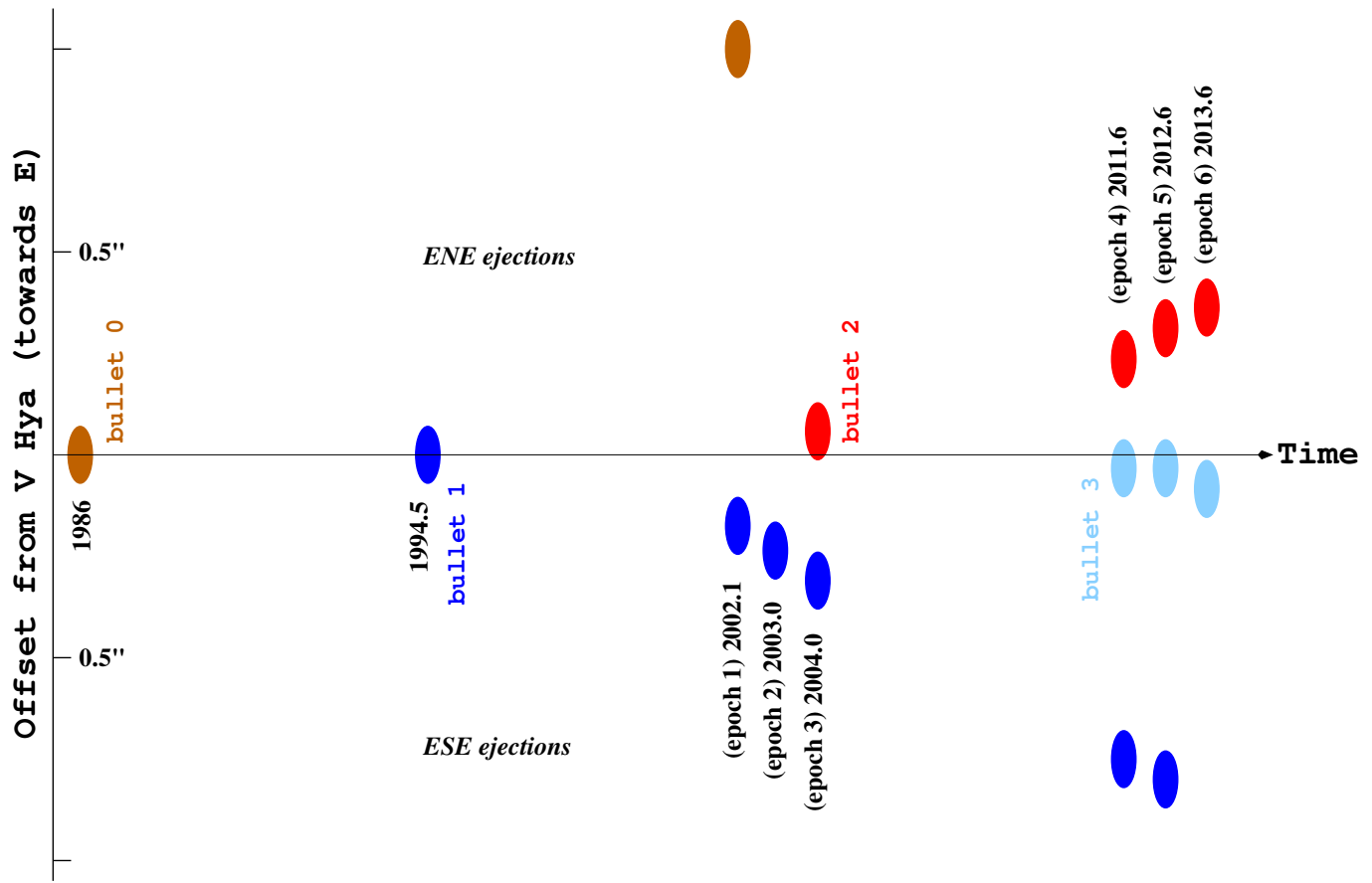}}
\caption{A schematic representation of the history of high-velocity bullet ejections from V\,Hya. The elliptical symbols show the location 
of these bullets, 
(which manifest themselves as on-source, detached and distant emission-line blobs at different epochs) relative to the center, with 
displacements (to scale) away from the x-axis in the vertical direction representing the offset (always E of the center) as 
seen in the slit that captures most of the blob emission. The placement of bullets that are not newly-ejected, in the top or bottom half 
of the plot depends on whether the 
corresponding detached or distant emission-line blob is primarily seen in the slit displaced N or S of the center, respectively: thus the 
top (bottom) half of the plot shows the bullets that were ejected to the ENE (ESE). The placement of newly-ejected bullets (which are 
seen primarily as the on-source blob in the central slit)  to the N or S of center, is assumed.
}
\label{blobhist}
\end{figure}

The on-source blob seen in epoch 3 (2004-01-12) represents bullet 2 (Fig.\,\ref{9800obs}a), which is also seen as 
the detached blob in epochs 4, 5, and 6 (Figs.\,\ref{go12227-bg}a,\,\ref{go12664-bg}a,\,\ref{go13053-bg}a). 
The distant emission blob seen in epoch 1 in slit $S_{+2b}$ (Fig.\,\ref{per1-per2-dist}a) likely 
represents an ejection that occurred $\sim8.5$\,yr prior to the one that was ejected during $1993.5-1995.5$, i.e. in $\sim$1986. We 
label this ejection bullet 0. The on-source blob seen in Period 2 (Figs.\,\ref{go12227-bg}b,\,\ref{go12664-bg}b,\,\ref{go13053-bg}b) represents the most recent ejection, i.e., bullet 3.
A schematic history of bullet ejections is shown in Fig.\,\ref{blobhist}.

\subsubsection{Emergence and Movement of a Newly-Ejected Bullet}\label{newblob}
We examine the emergence and early movement of bullet 3 by an inspection of 
the on-source blob emission during epochs $4-6$ (Figs.\,\ref{go12227-bg}b,\ref{go12664-bg}b,\ref{go13053-bg}b). During these epochs, 
we see a change in the blob position-velocity structure in the central slits, 
from one with no velocity gradient in epoch 4 to one with a clear gradient of about 200\,\kms\,arcsec$^{-1}$ in epoch 6 
(such that the more distant regions of the blob are blue-shifted relative to the less distant ones, implying differential acceleration 
across the bullet). 
The radial velocity of the peak emission in 
the blob also varies, increasing (in magnitude) from about $-160$\,\kms~in epoch 4 to $-200$\,\kms~in epoch 5, then decreasing to 
$-180$\,\kms~in epoch 6 (Table\,\ref{obssii}). 

\begin{figure}[htb]
\resizebox{0.5\textwidth}{!}{\includegraphics{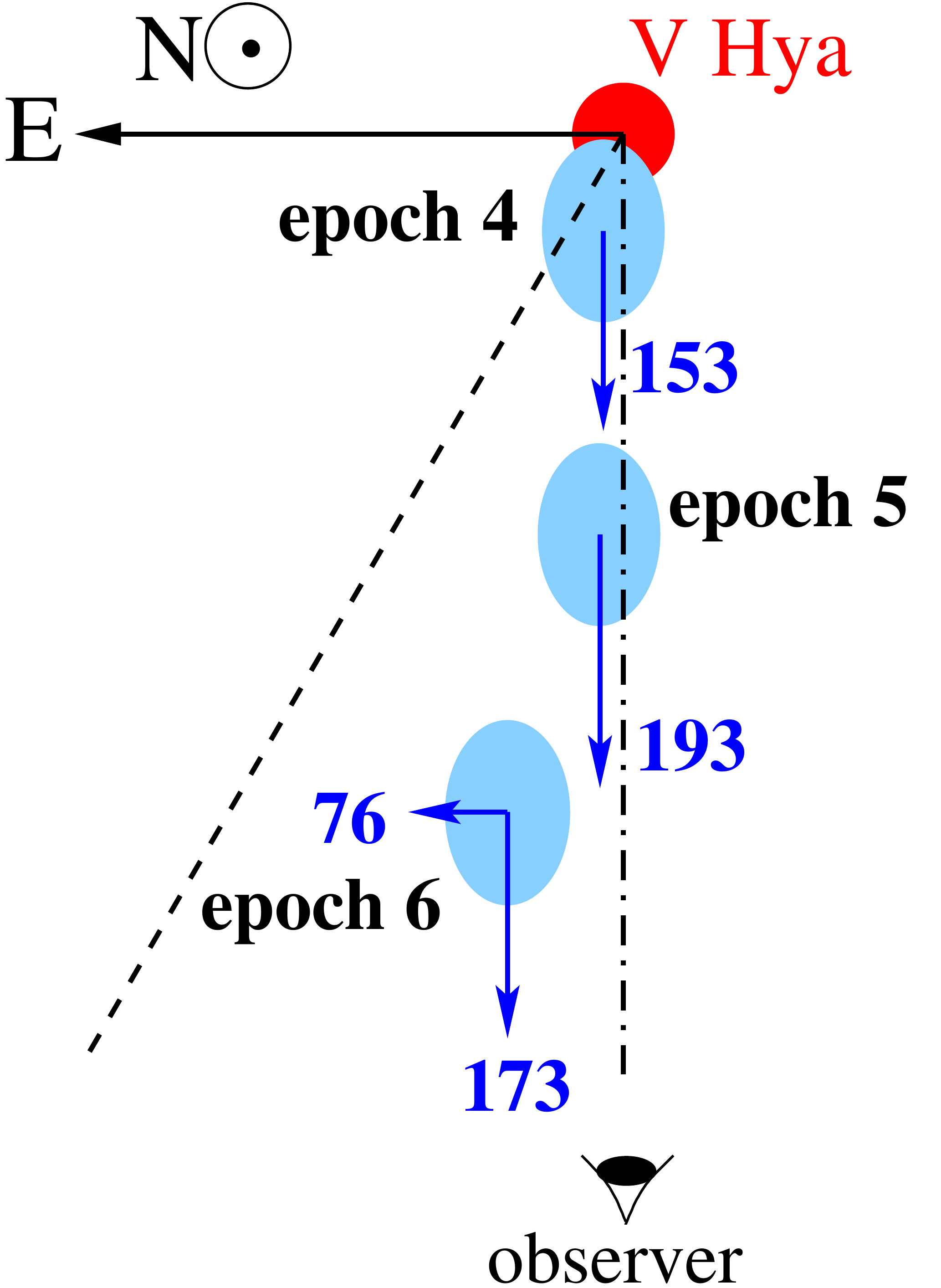}}
\caption{A schematic representation of the history of the bullet 3 ejection from V\,Hya. The elliptical symbols show the  
bullet locations during epochs 4--6. The arrows represent radial and tangential velocity vectors, roughly to scale (the numbers are the 
associated 
velocities in $\kms$, relative to V\,Hya's systemic velocity, $V_{hel}=-7$\,\kms). The dashed line shows the overall symmetry-axis 
of V\,Hya's extended high-velocity bipolar outflow. E is to the left, and N is out of the page (as shown).}
\label{bullet3move}
\end{figure}

Since there is no proper motion during epochs 4--5, we infer that bullet 3 moved roughly towards us 
along the line-of-sight during most of the time between epochs 4 and 5, accelerating at a rate of about 
40\,\kms\,yr$^{-1}$. Combining the proper 
motion between epochs 5--6 (which implies a tangential velocity, $V_t=76$\,\kms), with 
the radial velocity\footnote{\label{rvsys}relative to the systemic radial velocity of V\,Hya, $-7$\,\kms} in epoch 6, 
$V_r=-173$\,\kms, we find that (a) the bullet's motion tilted away from the line-of-sight (and thus towards the overall symmetry-axis 
of V\,Hya's extended high-velocity bipolar outflow, see Fig\,4 of Hirano et al. 2004) by about 23\arcdeg, sometime between epochs 5 and 6,  
and that (b) the 3-D outflow velocity in epoch 6 was $-189$\,\kms, which is  
almost the same as that\footref{rvsys} measured during epoch 5 ($-193$\,\kms). Thus there is little or no change in the speed of the bullet 
during epochs 5 and 6. A schematic of bullet 3's movement is shown in Fig.\,\ref{bullet3move}. 
We defer the investigation of possible mechanisms for the dynamical evolution of the bullet (e.g., a latitudinal density gradient in the 
ambient circumstellar medium in which the bullet is propagating, or a magnetic field) to a following 
paper (Scibelli, Sahai, \& Morris 2016, in preparation).

For bullet 2, we  
find that its radial velocity in epoch 3 ($-139$\,\kms) is significantly lower (in magnitude) than when it is next seen  
as a detached blob in epoch 4 ($-180$\,\kms). Although we cannot reconstruct the 3D motion of bullet 2 soon after ejection as 
we have done above for bullet 3, if we assume these motions are similar, then the change in radial velocity for bullet 2 from epoch 3 to 4  
could partly be due to a period of acceleration after ejection, and partly due to a change in inclination angle.

The size of the on-source blob (bullet 3) changes from epoch 4 to 6; during epoch 4, in addition to the emission seen in the central slit, 
weak [SII] emission is seen in 
slit $S_{+2b}$, but not in $S_{-2b}$ from this blob. During epoch 5, the blob is seen in all 3 slit locations (Fig.\,\ref{go12664-bg}), 
indicating that its size has increased and its extent has become resolved. Assuming a Gaussian 
cross-section for it in the N-S direction, the ratio of the average of the peak intensities 
seen in the $S_{+2b}$ and $S_{-2b}$ slits to that in the center slit (0.025), implies that its FWHM size is $0\farcs17$. 
During epoch 6, no [SII] emission can be detected from the on-source blob in the off-center slits -- 
the ratio of the 3$\sigma$ upper limit for the off-center intensity to 
the intensity in the central slit is about 0.007. This implies that the (projected) emitting extent of bullet 3, in the N-S direction, 
decreases from epoch 5 to 6. Along the slit direction, i.e. E-W, the 
half-power size is $0\farcs12$, $0\farcs13$, and $0\farcs11$, in epochs 4, 5 and 6, which is very similar to the FWHM size of the 
continuum source measured in the on-source slits ($0\farcs13$). Making the reasonable assumption that the continuum source is unresolved, 
we conclude that emission from bullet 3 is also unresolved along the E-W direction. Its deconvolved size 
along the N-S direction in epoch 5 is about $0\farcs11$, or 44\,AU at D=400 pc.



\subsubsection{A Flip-Flop Phenomenon for Bullet Ejections}\label{flipflop}
There are qualitative and quantitative differences between the bullets in periods 1 and 2. 
Comparing the detached and distant blobs in these periods, we see an interesting pattern related to their locations -- 
(a) in Period 1, the detached blob (bullet 1) is seen predominantly in the slit immediately south of center (i.e., $S_{-2b}$ or $S_{-1t}$), whereas 
in Period 2, the detached blob (bullet 2) is seen predominantly in the slit north of center (i.e., $S_{+2b}$), and (b) in Period 1, the distant 
blob (bullet 0) is seen dominantly in slit $S_{+2b}$, whereas in Period 2, the distant blob (bullet 1) 
is seen dominantly in slit $S_{-2b}$. This implies that  
the bullet-ejection axis flips periodically between the east-southeast (ESE) and east-northeast (ENE) directions, 
in the plane of the sky. This flip-flop pattern is represented schematically in Fig.\,\ref{blobhist}. 

In addition, the observed radial-velocities of the blobs show that the bullet-ejection axis also flip-flops in a plane that is 
perpendicular to the sky-plane. First, 
the radial velocities of the on-source blobs in Period 2 (bullet 3) are significantly more blue-shifted than that of the on-source blob  
in Period 1 (bullet 2). Second, the radial velocities of the detached blobs in Period 2 (bullet 2) are significantly less blue-shifted compared to 
those of the detached blobs in Period 1 (bullet 1). Third, the radial velocity of the 
distant blobs in Period 2 (bullet 1) appears to be more blue-shifted than that of the distant blob in Period 1 (bullet 0).
All of these observational features associated with the flip-flop phenomenon are summarized in 
Table\,\ref{tabflipflop}.

Interestingly, ground-based observations of the [SII]$\lambda4069.7$ line obtained during 
1986\,Dec--1990\,May by Lloyd Evans (1991) show a radial velocity of $-166$\,\kms, consistent with the less blue-shifted 
radial velocity of bullet 0 (ejected around 1986). Due to the typical $\sim1{''}$ ground-based seeing, these data would 
also have included emission from the the higher blue-shifted radial-velocity bullet expected to have been ejected in 1977.5, but 
its contribution would be relatively minor as it would be much fainter than the more recently-ejected bullet, given 
that the blob emission intensity decreases rapidly once the bullet has moved far away from the central source. 

Other systematic differences between blobs in different periods include the following: 
(a) the on-source blob in Period 1 is clearly 
seen only in one epoch (epoch 3), whereas in Period 2 it is seen in all three epochs (epochs 4, 5, \& 6),  
(b) the blue-shift of the detached blob is about 25\% higher in Period 1 compared 
to Period 2, and (c) the average brightness of the detached blob is about a factor $5$ higher in Period 1 compared 
to Period 2.

We also find that the 
on-source blob in epoch 3 shows a velocity-gradient across its length, with a blue-shift that decreases with spatial offset 
from the center. In contrast, the on-source blob in Period 2 does not show a velocity gradient in epoch 4, and two years 
later, when it shows a clear velocity gradient, it is in the opposite sense to that in Period 1, i.e., the blue-shift increases 
with spatial offset from the center.

\subsection{Physical Conditions in the Bullets}\label{blobphys}
The STIS specta, covering the wavelength range 4050-4359\AA, include not only the 
[SII] doublet at 4069.75 and 4077.50\AA, but also several emission lines from [FeII]. The 
most intense of these is the 4245.16\AA~(E2, $1872.57-25428.78$\,cm$^{-1}$) line
followed by the 4288.6\AA~(E2, $0.0-23317.63$\,cm$^{-1}$) line; the fractional intensity of the former is about 0.1--0.2 
times that of the [SII]$\lambda4069.7$ line. 


We have used the CLOUDY code (version 13.03, last described by Ferland et al. 2013) to 
constrain the density and temperature in the detached blob in Period 1, and the on-source blob in Period 2, 
by fitting specific emission-line ratios 
for these lines, assuming a collisionally-ionized medium. We are not able to carry out such an analysis for other blobs because 
of the low S/N of lines other than the [SII]$\lambda4069.7$ line. 

\subsubsection{Bullet 1: Detached Blob in Period 1}
We find that in order to reproduce the observed [FeII]$\lambda$4245/[SII]$\lambda$4069.7 
or [FeII]$\lambda$4288/[SII]$\lambda$4069.7 line ratios, the kinetic temperature, 
$T_{kin}\sim1.1\times10^4$\,K; for higher (lower) temperatures the model 
ratios are significantly lower (higher) than observed. The temperature sensitivity of these line ratios   
results from the relatively large changes that occur in the fractional ionization of S at 
$\sim10^4$\,K -- at such temperatures, most of the S is in atomic form and small changes 
in $T_{kin}$ make large changes in the fraction  of S$^+$, whereas the fractional 
ionization of Fe is much less affected. 

\begin{figure}[htb]
\resizebox{0.5\textwidth}{!}{\includegraphics{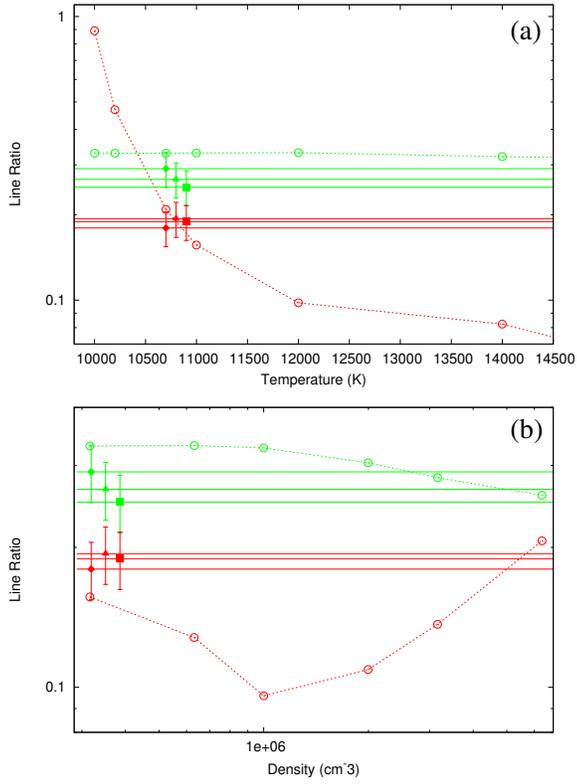}}
\caption{Observed and model line ratios for the detached blobs, for 3 epochs in Period 1, as a function of (a) temperature, (b) density. 
Green solid lines represent the observed [SII]\,$\lambda$4077.5/$\lambda$4069.7 ratios, 
and red solid lines represent the observed $0.5\times$([FeII]$\lambda$4245+$\lambda$4285)/[SII]\,$\lambda$4069.7 line ratios. 
The observed-ratio lines are tagged for datasets GO\,9100, 9632 and 9800 with a filled diamond, triangle and 
square symbol (and error bar), respectively, near a temperature of about $1.08\times10^4$\,K in panel $a$, 
and near a density of about $3.5\times10^5$\,cm$^{-3}$ in panel $b$. 
The model ratios, shown as open circular symbols joined by dotted lines, are derived using CLOUDY, for a density of 
$3.2\times10^5$\,cm$^{-3}$ in panel $a$ and for a temperature of 11,000\,K in panel $b$. }
\label{lineratios1}
\end{figure}

The relative intensity of the [SII] doublet, 
$\lambda$4077.5/$\lambda$4069.7 (hereafter [SII] doublet ratio) is rather insensitive to temperature. In 
Fig.\,\ref{lineratios1}a, we show a plot of the observed and model line-ratios for a range of 
temperatures (at a density of $3.2\times10^5$\,cm$^{-3}$); the data are reasonably well fitted 
for $T_{kin}\sim11,000$\,K. The two 
[FeII] lines at 4245 and 4288\,\AA~vary similarly with temperature over a wide range. Hence, we have used the 
$0.5\times$([FeII]$\lambda$4245+$\lambda$4285)/SII\,$\lambda$4069.7 line ratio (hereafter [FeII]/[SII] line 
ratio) for comparison in these plots, since the [FeII] lines are relatively weak.

Both S and Fe may be partially incorporated into dust grains, thus affecting the [FeII]/[SII]  
line ratio discussed above, but it is likely that much of the dust is destroyed 
in the violent interaction of the high-speed bullet with a more slowly moving ambient 
circumstellar medium. But if some fraction of each of these elements does remain 
locked up in dust grains, the net effect is to decrease the [FeII]/[SII] line ratio, 
since Fe is less volatile than S. Under these circumstances, the observed 
[FeII]/[SII] line ratio represents a lower limit for the models (which use the cosmic S and Fe abundances), and  
$T_{kin}<11,000$\,K. 

We note that the observed [SII] doublet ratio appears somewhat lower than the model ones at all temperatures in Fig.\,\ref{lineratios1}a.  
We therefore investigate the dependence of the above line ratios on density for a temperature of 11,000\,K, 
and find that the [SII] doublet ratio decreases with increasing density (Figs.\,\ref{lineratios1}b). 
The [FeII]/[SII] ratio goes through a 
minimum at a density of $\sim10^6$\,cm$^{-3}$, and then increases again. A reasonable fit to both of the ratios 
implies a density of $\sim5\times10^6$\,cm$^{-3}$. A rough estimate of the mass of bullet 1, $M_{b}$, can then be made assuming it to  
be a sphere with uniform density and physical size corresponding to the deconvolved half-intensity angular extent of the detached blob as
measured along the slit, in slit $S_{-2b}$ in epoch 1 (about $0\farcs1$ or 40 AU)\footnote{Detailed spatio-kinematic modelling to derive the blob geometry and density 
structure are deferred to a following paper, Scibelli, Sahai, \& Morris (2016, in preparation)}; we find $M_{b}\sim10^{27}$\,g.  
For bullet 2, we estimate a mass of about 0.7 times that of bullet 1, 
using the fractional [SII]$\lambda$4069.7 flux (0.2) and deconvolved size (1.35) of the detached blob in Period 2 (epoch 4), 
relative to that in Period 1 (epoch 1), and assuming that the line emissivity varies as density-squared. 

Hence the average mass-loss rate via bullet ejection in V\,Hya is 
about $4.7\times10^{-8}$\,\my. The momentum flux (mass-loss rate $\times$ outflow-speed) in the bullets is 
roughly comparable to that in the complex high-velocity molecular outflow seen towards V\,Hya using 4.6\,\micron~CO spectroscopy 
(Sahai, Sugerman \& Hinkle 2009) -- the latter have a 
mass-loss rate ($\sim10^{-7}$\,\my)\footnote{There is a typo in 
the exponent of the value of $\omega$ in $\S$5.1.1 of Sahai, Sugerman \& Hinkle (2009); the correct value is $\omega=1.2\times10^{-2}$, implying $\mdot_{HV} > 10^{-7}$\,\my} about a factor 2 higher, and an average outflow velocity ($\sim100$\,\kms) a factor 2 lower than the bullets. 
This it seems likely that these high-speed molecular outflows seen in the IR in V\,Hya are driven by the bullet ejections, 
via a momentum-conserving hydrodynamic interaction of the latter with V\,Hya's slowly expanding AGB wind.


\subsubsection{Bullet 3: On-Source Blob in Period 2}
\begin{figure}[htb]
\resizebox{0.5\textwidth}{!}{\includegraphics{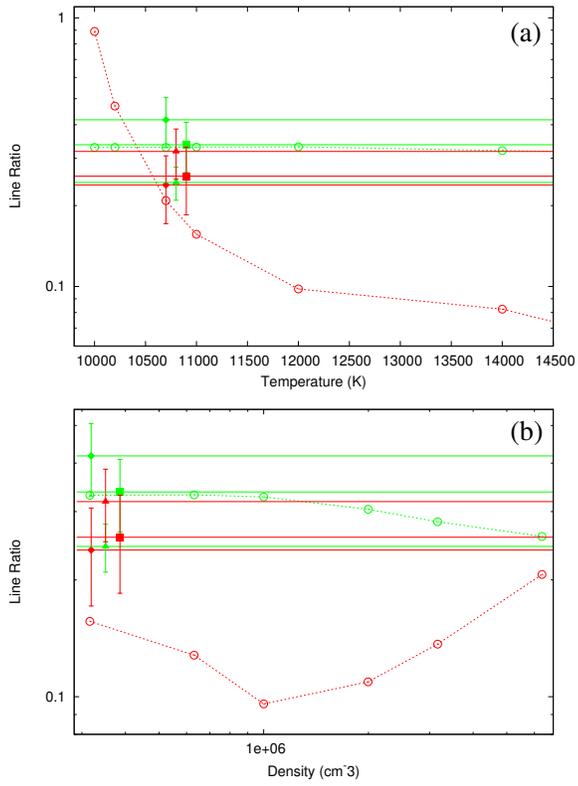}}
\caption{As in Fig.\,\ref{lineratios1}, but for the on-source blob, for 3 epochs in Period 2 (datasets GO\,12227, 12664, 13053).}
\label{lineratios2}
\end{figure}

The observed relative line intensities for the on-source blob are shown in Fig.\,\ref{lineratios2}, together with model ratios. 
Due to the presence of the strong stellar continuum superimposed on the on-source blob emission, the line intensities and ratios are more 
uncertain than in the case of the detached blob (discussed above). On average, both the [SII] doublet and [FeII]/[SII] line ratios
for the on-source blob in Period 2 appear to be somewhat higher than for the detached blob in Period 1.  
We are not able to simultaneously fit both these line ratios; the biggest discrepancy is for epoch 5, where the S/N is highest -- here the 
[SII] doublet ([FeII]/[SII] line) ratio is signficantly lower (higher) than the model value. A plausible physical reason for this discrepancy 
may be the blob material being far from collisional equilibrium during its early ejection history.

\subsection{A Binary Model for Blob Ejection in V\,Hya}\label{binmodel}
We propose a binary model for the $\sim8.5$\,yr period of bullet ejection and the flip-flop phenomenon in V\,Hya's bullet ejection history. In this model, 
the bullet ejection is associated with the periastron passage of a binary companion in an eccentric orbit with an orbital period of $\sim8.5$\,yr. 
The bullets are ejected from an accretion disk formed around the companion that results from the gravitational capture of matter from the primary wind, 
and perhaps directly from the pulsating atmosphere. 
The amount of matter ejected by an accretion disk is, in general, a few percent of the accretion rate (Livio 1997). We expect   
the accretion rate to increase sharply at periastron because of two reasons. First, the secondary is then moving through, or near, 
the extended atmosphere of the primary, or the high-density base 
of the primary's wind. Second, the tidal effects on the accretion disk are greatest at this point, which could increase the outward 
transfer of angular momentum in the disk, thus bringing material efficiently into the inner-disk jet-launching zone. 
Thus the ejection of the high-speed bullets during periastron passage is a reasonable expectation.

On both theoretical and empirical grounds, the speed
of a jet driven from an accretion disk is expected to be of the 
order of the Keplerian velocity close to the central accreting
object (Livio 1997), $V_{Kep}=438\,\kms\,(M_c/R_c)^{0.5}$, where $M_c$ and $R_c$ are the companion mass and radius in solar units. 
Thus the $\sim200-250$\,\kms~speeds that we find for the high-speed blob imply a solar or sub-solar 
main-sequence (MS) companion. 
We find, for a range of nominal primary masses ($1 \lesssim M_p (\ms)\lesssim 2$) and companion masses ($M_c < M_p$), that the eccentricity 
has to be relatively large, $e \gtrsim 0.6$, in order for the companion to approach the primary within the latter's stellar envelope at 
periastron, 
assuming a primary radius $3.8\times10^{13}\,(D/500 pc)$ (Knapp et al. 1999), or 2\,AU at D=400\,pc.

We note that knotty jet-like outflows have been found to exhibit a ``flip-flop" behaviour in the symbiotic Mira R \,Aqr, 
which may be related to what we see in 
V\,Hya. In R\,Aqr, a one-sided jet apparently flips direction from  
one side of the disk to the other every $\sim17$\,yr, presumably corresponding to periastron passage in a binary with the same 
period (Nichols et al. 2007, Stute \& Sahai 2007). 
The high-speed outflows in R Aqr and V Hya can be understood within a common framework if one assumes that in both cases, (a) 
the outflows are bipolar but stronger on opposite sides of the disk in successive periastron passages, and (b) 
there is also a change in the orientation of the axis of ejection. 
We know that the high-speed outflow in V Hya on large scales is bipolar from the mm-wave CO mapping (Kahane et al. 1996), Hirano et al. 2004) -- 
it is therefore likely that the western side of the optical high-speed blobby outflow is not seen in the STIS data because 
it remains obscured behind V\,Hya's equatorial torus\footnote{this structure is to be distinguished from the much smaller disk around the companion} at all times.  
 
A planar disk would emit jets symmetrically and simultaneously along its symmetry axis, especially if 
the magnetic axis of the companion was aligned with the disk axis. 
However, there are at least a couple of ways in 
which such symmetry may be broken, enabling the flip-flop phenomenon observed in V\,Hya -- we discuss these below. 

(i) The disk is warped. Self-warping of the disk may occur as a result of the strong radiation field 
from the primary: Pringle (1996) has investigated such warping for disks with a central radiation source. In the case of V\,Hya, 
radiation-induced warping would be dominant during periastron passage when the companion and disk would be closest to the primary. 
Warping could also occur as a result of the stronger interaction of the side of the 
disk that is closer to the primary compared to the one that is more distant. This interaction consists of 
a tidal interaction and a direct physical interaction (i.e., frictional drag) with the atmospheric medium of the primary, which 
combined with a radial density gradient in the atmosphere, could produce a differential torque that could warp the disk.
The warped disk would exhibit precession, and in order to account for the flip-flopping behavior of the bullet-ejection axis, 
this precession would need to have a period twice the orbital period.

(ii) The disk-axis flips back-and-forth due to a short-duration torque during every periastron passage, as a result of the 
differential drag across the disk as it moves through the outer atmosphere of the primary, or the high-density base 
of its wind.


\subsubsection{Comparison with Previous Observations of Binarity}

V\,Hya's optical spectrum shows anomalously broad lines with a width that has a mean value of 13.5\,\kms~and shows a 
total variation of 9\,\kms, in concert 
with the photometric phase of the primary's 529 day pulsation period, consistent with the expectation for a rapidly rotating star that 
conserves angular momentum as it rotates (Barnbaum, Morris, \& Kahane 1995). These authors propose that the inferred fast rotation of the outer layers of V\,Hya's 
stellar atmosphere is due to a binary companion in a common-envelope configuration with the primary (although alternative broadening 
mechanisms are possible, these authors consider them less likely.) In our model, the dominant transfer of angular momentum to V\,Hya's outer 
layers would occur during periastron passage.

V\,Hya also shows a very long period, ($6160\pm400$\,d) in its light curve (Knapp et al. 1999, hereafter K99) 
when the star undergoes deep minima. K99 suggest that V\,Hya is an eclipsing 
binary with an orbital period of 6160\,d, and 
the periodic deep minima that occur result from enhanced extinction by a dust-cloud associated with the
companion. The $6160\pm400$\,d period, at the lower end of the range, is 16.8\,yr, very similar to period for the flip-flop, estimated to be twice the 
bullet ejection period, i.e., $\sim$17 yr . 
So it is plausible that these periods are really the same, as it would be rather unlikely to have two such similar periods associated with 
different physical mechanisms. If so, then Period 1 (Period 2) of our STIS observations coincides with phase 0 (0.5), i.e. 
the maxima (minima) of the long-period light-curve.

We propose that the deep minima in the long-period light-curve are associated with extinction associated with the bullets, and 
that this happens every alternate 8.5\,yr orbital cycle as follows. We have shown earlier (\S\,\ref{flipflop}) that in addition to 
the flip-flop of the bullet-ejection axis in the sky-plane (causing the ejection direction to move between ESE and ENE), there is 
also a corresponding flip-flop in a plane perpendicular to 
the sky-plane (causing the pronounced jumps in radial velocity between successive bullets). 
We suggest that it is this line-of-sight component of the flip-flop that 
causes the bullets to be ejected in such a way that as they expand and move across our line-of-sight to V\,Hya, 
they alternate between being either in front of the primary star or behind it. 
In this scenario, the bullet ejected in Period 1 (i.e., bullet 2) lies behind the primary star during epochs 1--2, and 
only emerges in epoch 3; the bullet-ejection occurs most likely somewhere between epoch 1 and epoch 3. 

As noted earlier, V\,Hya lacks X-ray emission, even though its relatively high FUV/NUV flux ratio (a 
strong indicator of X-ray emission in fuvAGB stars) indicated an expected X-ray luminosity 
more than two orders of magnitude higher than the observed $3\sigma$ upper limit (see Fig.\,4b, Sahai et al. 2015). 
Our eccentric-orbit model, in 
which the accretion rate decreases steeply away from periastron, provides a natural explanation for V\,Hya's 
lack of X-ray emission, if we make the reasonable assumption that the latter is 
related to the accretion rate.  
The X-ray observations were made on 2013-12-18 and thus at least 2.5\,yr after periastron passage, 
when the accretion rate must have dropped sharply 
from its value at periastron.

The map of CO J=2--1 emission from V\,Hya (Fig.\,3 of Hirano et al. 2004) shows high-speed 
emission blobs that are modestly offset to the N and S from an overall E--W axis in a staggered pattern that is qualitatively 
consistent with the flip-flop phenomenon seen in our data. The expansion time-scale of these blobs is $\sim300$\,yr, derived by dividing their  
radio distance from the center ($\sim10{''}\,tan\,(90\arcdeg-i)$ at 400 pc or 7000\,AU, taking Hirano et al.'s value for the 
angle between the outflow direction and the line-of-sight, $i=30\arcdeg$) by their mean radial velocity (110\,\kms). Thus, it appears  
that the precessional-resonance of the accretion disk that is responsible for the flip-flop phenomenon in V\,Hya lasts for 
at least several hundred years.

\section{Concluding Remarks}\label{conclude}
Our study provides an unprecedented and detailed view of the history and characteristics of bullet-like high-speed ejections from the 
carbon star, V\,Hya. The main observational features of this ejection in V\,Hya are well explained by a binary model in which the bullet ejection is 
associated with the periastron passage of a main-sequence binary companion in an eccentric orbit with an orbital period of $\sim8.5$\,yr, 
and in which the bullet-ejection direction flip-flops around an E-W axis (with discrete and repeatable components 
to the angular displacements both in the sky-plane, and perpendicular to it) at twice the orbital 
period. 

Our model can be used to predict the locations of previously ejected bullets in V\,Hya and 
future epochs at which new bullets will emerge. Thus, we expect that bullet 3, seen emerging in Period 2, will move towards the ESE, 
and a new one (bullet 4) will 
be ejected in 2020 towards the ENE. New multi-epoch observations with STIS can test these predictions.

The high-speed bullet-like ejections that we have found in V\,Hya may are apparently a relatively common phenomenon during the transition 
from the AGB to the PN evolutionary phase, as they 
have been found in several PPNe (e.g., Hen\,3-1475: Borkowski et al. 1997; IRAS22036+5306: Sahai et al. 2006; CRL\,618: Lee et al. 2013, Balick et al. 2013) and PNe 
(e.g., Fleming\,1: Lopez et al. 1993, KjPn\,8: Lopez et al. 1995; MyCn18: O'Connor et al. 2000, Hen\,2-90: Sahai \& Nyman 2000) as well. AGB stars with a binary 
companion in a relatively high-eccentricity orbit, as we propose for V\,Hya, provide a natural mechanism to enhance the accretion rate every periastron passage, 
leading to strongly variable and highly-collimated mass-loss, i.e., or what some have termed knotty jets. In our eccentric binary model for bullet ejection, knotty jets that 
are linearly aligned along a radial vector, would emanate from systems in which  
the accretion disk around the companion does not undergo significant precession. On the other hand, 
non-resonant disk precession\footnote{i.e., the ratio of orbital period to the disk precession period is not expressible as a 
ratio of small integers} 
would generate objects showing multiple, diametrically-opposed knot-pairs around the central object, 
with successive knots having progressively varying position angles 
(e.g., as in He\,3-1475, Fleming\,1, KjPn\,8, \& MyCn18)\footnote{see Cliffe et al. (1995) for a numerical hydrodynamical simulation of such a jet}.

Furthermore, if, within the context of the ``precessional-resonance" case of the model that we propose for V\,Hya, 
the change in the disk-axis as a result of precession is 
sizeable, and the system ``remembers" the different, discrete, orientations of this axis over relatively long timescales, it could explain the 
formation of multipolar PNe. Thus, the Starfish PNe (He\,2-57 \& M\,1-37: Sahai 2000), with three pairs of bipolar lobes,
would correspond to a 3:1 precessional-orbital resonance, while quadrupolar PNe  -- a subclass of multipolar PNe, that have two 
pairs of bipolar lobes with differing orientations (e.g., M\,2-46, K\,3-24, \& M\,1-75: Manchado et al. 1996) -- would correspond to a  
2:1 precessional-orbital resonance as in V\,Hya. Theoretical 
investigations are warranted to explore the precessional-resonance model for V\,Hya, as 
well as the implications that this model may have for multipolar systems.

\acknowledgements RS's contribution to the research described here was carried out at the Jet Propulsion Laboratory (JPL), 
California Institute of Technology, under a contract with NASA, with financial support was provided by NASA, in part from an STScI HST 
award (GO\,12227.01). SS's contribution was carried out during her tenure as a NASA Undergraduate Intern (UI) at JPL.


\clearpage

\scriptsize
\begin{turnpage}
\begin{longtable}{p{1.0in}p{0.3in}p{0.35in}p{0.35in}p{0.35in}p{0.35in}p{1.0in}p{0.4in}p{0.8in}}
\caption[]{Log of Observations}\\
\hline \\[-2ex]
   \multicolumn{1}{l}{\textbf{Epoch}} &
   \multicolumn{1}{l}{\textbf{Program}} &
   \multicolumn{1}{l}{\textbf{DataSetRoot}} &
   \multicolumn{5}{c}{\textbf{Slit}}  &
   \multicolumn{1}{l}{\textbf{ExpTime\footnotemark[1]}} \\[0.5ex]
   \multicolumn{1}{l}{\textbf{(\#)\,\,\,\,Date}} &
   \multicolumn{1}{l}{\textbf{GO}} &
   \multicolumn{1}{c}{\textbf{}} &  
   \multicolumn{1}{l}{\textbf{Name}} &
   \multicolumn{1}{l}{\textbf{Offset\footnotemark[2]}} &
   \multicolumn{1}{l}{\textbf{Width}} &  
   \multicolumn{1}{c}{\textbf{Dither\footnotemark[3]}} &
   \multicolumn{1}{l}{\textbf{PA($\arcdeg$)}} &
   \multicolumn{1}{l}{\textbf{(sec)}} \\[0.5ex] \hline
   \\[-1.8ex]
\endfirsthead

\multicolumn{3}{c}{{\tablename} \thetable{} -- Continued} \\[0.5ex]
\hline \\[-2ex]
   \multicolumn{1}{l}{\textbf{Epoch}} &
   \multicolumn{1}{l}{\textbf{Program}} &
   \multicolumn{1}{l}{\textbf{DataSetRoot}} &
   \multicolumn{5}{c}{\textbf{Slit}}  &
   \multicolumn{1}{l}{\textbf{ExpTime\footnotemark[1]}} \\[0.5ex]
   \multicolumn{1}{l}{\textbf{(\#)\,\,\,\,Date}} &
   \multicolumn{1}{l}{\textbf{GO}} &
   \multicolumn{1}{c}{\textbf{}} &  
   \multicolumn{1}{l}{\textbf{Name}} &
   \multicolumn{1}{l}{\textbf{Offset\footnotemark[2]}} &
   \multicolumn{1}{l}{\textbf{Width}} &  
   \multicolumn{1}{c}{\textbf{Dither\footnotemark[3]}} &
   \multicolumn{1}{l}{\textbf{PA($\arcdeg$)}} &
   \multicolumn{1}{l}{\textbf{(sec)}} \\[0.5ex] \hline
   \\[-1.8ex]
\endhead
\multicolumn{1}{l}{\textbf{}} &
\multicolumn{1}{l}{\textbf{}} &
\multicolumn{1}{c}{\textbf{Period 1}} \\[0.5ex] \hline
\\[-1.8ex] 
(1) 2002-01-28  & 9100  &  o6fn01010,20    & $S_{-2b}$ & $-0\farcs2$ &   $0\farcs2$    &   $0.0$, $0\farcs025$  & 90.06 &  $720$, $858$  \\
            &       &  o6fn01030,40    & $S_{0b}$ & $0.0$        &   $0\farcs2$    &   $0.0$, $0\farcs025$  & " &  $792$, $720$  \\
            &       &  o6fn01050,60    & $S_{+2b}$ & $0\farcs2$  &   $0\farcs2$    &   $0.0$, $0\farcs025$  & " & $720$, $692$   \\
(2) 2002-12-29  & 9632  &  o8hs01010,20,30 & $S_{-2t}$ & $-0\farcs2$ &   $0\farcs1$    &  $-0\farcs225$, $0.0$, $0\farcs225$ & 87.06 &  $402$, $402$, $402$  \\
            &       &  o8hs01040,50,60 & $S_{-1t}$ & $-0\farcs1$ &   $0\farcs1$    &   $-0\farcs225$, $0.0$, $0\farcs225$  & " & $402$, $375$, $402$ \\
            &       &  o8hs01070,80,90 & $S_{0t}$ & $0.0$        &   $0\farcs1$    &  $-0\farcs225$, $0.0$, $0\farcs225$  & " &  $402$, $402$, $402$ \\
            &       &  o8hs010a0,b0,c0 & $S_{+1t}$ & $0\farcs1$  &   $0\farcs1$    &  $-0\farcs225$, $0.0$, $0\farcs225$   & " & $442$, $402$, $402$ \\
            &       &  o8hs010d0,e0,f0 & $S_{+2t}$ & $0\farcs2$  &   $0\farcs1$    &  $-0\farcs225$, $0.0$, $0\farcs225$   & " & $402$, $402$, $402$ \\
(3) 2004-01-12  & 9800  &  o8or01010,20,30 & $S_{-1t}$ & $-0\farcs1$ &  $0\farcs1$     &  $-0\farcs225$, $0.0$, $0\farcs225$  & 87.06  & $522$, $522$, $522$ \\
            &       &  o8or01040,50,60 & $S_{0t}$ & $0.0$        &  $0\farcs1$     &  $-0\farcs225$, $0.0$, $0\farcs225$  &  " & $522$, $522$, $522$ \\
            &       &  o8or01070,80,90 & $S_{+1t}$ & $+0\farcs1$ &  $0\farcs1$     &  $-0\farcs225$, $0.0$, $0\farcs225$  &  " & $522$, $522$, $522$ \\
            &       &  o8or010a0,b0,c0 & $S_{+2t}$ & $+0\farcs2$ &  $0\farcs1$     &  $-0\farcs225$, $0.0$, $0\farcs225$   & " & $522$, $522$, $522$ \\
\hline \\[-2ex]
\multicolumn{1}{l}{\textbf{}} &
\multicolumn{1}{l}{\textbf{}} &
\multicolumn{1}{c}{\textbf{Period 2}} \\[0.5ex] \hline
\\[-1.8ex]
(4) 2011-07-07  & 12227 &  obkg01010,20,30 & $S_{+2b}$ & $+0\farcs2$ &  $0\farcs2$     &  $0.0$, $0\farcs2285$, $0\farcs4570$  & 83.56 & $898$, $898$, $738$ \\
            &       &  obkg01040,50,60 & $S_{0b}$ & $0.0$        &  $0\farcs2$     &   $0.0$, $0\farcs2285$, $0\farcs4570$  & " & $738$, $738$, $738$ \\
            &       &  obkg01070,80,90 & $S_{-2b}$ & $-0\farcs2$ &  $0\farcs2$     &   $0.0$, $0\farcs2285$, $0\farcs4570$  & " & $738$, $738$, $738$ \\
(5) 2012-07-14  & 12664 &  obr101010,20,30 & $S_{+2b}$ & $+0\farcs2$ &  $0\farcs2$     &  $0.0$, $0\farcs3803$, $0\farcs7607$  & 83.75 & $898$, $898$, $738$ \\
            &       &  obr101040,50,60 & $S_{0b}$ & $0.0$        &  $0\farcs2$     &   $0.0$, $0\farcs3803$, $0\farcs7607$  & " & $738$, $738$, $738$ \\
            &       &  obr101070,80,90 & $S_{-2b}$ & $-0\farcs2$ &  $0\farcs2$     &   $0.0$, $0\farcs3803$, $0\farcs7607$  & " & $738$, $738$, $738$ \\
(6) 2013-07-17  & 13053 &  oc3601010,20,30 & $S_{+2b}$ & $+0\farcs2$ &  $0\farcs2$     &  $0.0$, $0\farcs3803$, $0\farcs7607$ & 85.27 & $914$, $914$, $728$ \\
            &       &  oc3601040,50,60 & $S_{0b}$ & $0.0$  	 &  $0\farcs2$     &   $0.0$, $0\farcs3803$, $0\farcs7607$  & " &  $728$, $728$, $728$ \\
            &       &  oc3601070,80,90 & $S_{-2b}$ & $-0\farcs2$ &  $0\farcs2$     &   $0.0$, $0\farcs3803$, $0\farcs7607$  & " & $728$, $728$, $728$ \\
\label{obslog}
\end{longtable}
\end{turnpage}
\footnotetext[1]{Exposure Time at each dither position successively}
\footnotetext[2]{Mosaic Offset along a direction orthogonal to the slit}
\footnotetext[3]{Dither Offset along the slit direction}




\clearpage
\scriptsize
\begin{turnpage}
\begin{longtable}{p{0.8in}p{0.5in}p{0.35in}p{0.35in}p{0.35in}p{0.5in}p{0.35in}p{0.35in}p{0.35in}p{0.5in}p{0.5in}}
\caption[]{Observational Properties Ia -- Detached \& On-Source Blobs ([SII] $\lambda$4069.75, Continuum)}\\
\hline \\[-2ex]
   \multicolumn{1}{l}{\textbf{Epoch}} &
   \multicolumn{1}{l}{\textbf{Slit}} &
   \multicolumn{4}{c}{\textbf{Detached}} &
   \multicolumn{4}{c}{\textbf{On-Source}} &
   \multicolumn{1}{l}{\textbf{Continuum}} \\[0.5ex]
   \multicolumn{1}{l}{\textbf{(\#) Date}} &
   \multicolumn{1}{l}{\textbf{Name}} &
   \multicolumn{1}{l}{\textbf{Offset}} &
   \multicolumn{1}{l}{\textbf{$V_P$}} &
   \multicolumn{1}{l}{\textbf{fwhm}} &
   \multicolumn{1}{l}{\textbf{Int.\footnotemark[1]}} &  
   \multicolumn{1}{l}{\textbf{Offset}} &
   \multicolumn{1}{l}{\textbf{$V_P$}} &
   \multicolumn{1}{l}{\textbf{fwhm}} &
   \multicolumn{1}{l}{\textbf{Int.}} & 
   \multicolumn{1}{l}{\textbf{Int.}} \\[0.8ex]
   \multicolumn{1}{l}{\textbf{}} &
   \multicolumn{1}{l}{\textbf{}} &
   \multicolumn{1}{l}{\textbf{arcsec}} &
   \multicolumn{1}{l}{\textbf{\,\kms}} &
   \multicolumn{1}{l}{\textbf{\,\kms}} &
   \multicolumn{1}{l}{\textbf{cgs}} &  
   \multicolumn{1}{l}{\textbf{arcsec}} &
   \multicolumn{1}{l}{\textbf{\,\kms}} &
   \multicolumn{1}{l}{\textbf{\,\kms}} &
   \multicolumn{1}{l}{\textbf{cgs}} & 
   \multicolumn{1}{l}{\textbf{cgs}}\\[0.5ex] \hline
   \\[-1.8ex]
\endfirsthead
\multicolumn{3}{c}{{\tablename} \thetable{} -- Continued} \\[0.5ex]
\hline \\[-2ex]
   \multicolumn{1}{l}{\textbf{Epoch}} &
   \multicolumn{1}{l}{\textbf{Slit}} &
   \multicolumn{4}{c}{\textbf{Detached}} &
   \multicolumn{4}{c}{\textbf{On-Source}} &
   \multicolumn{1}{l}{\textbf{Continuum}} \\[0.5ex] 
   \multicolumn{1}{l}{\textbf{(\#) Date}} &
   \multicolumn{1}{l}{\textbf{Name}} &
   \multicolumn{1}{l}{\textbf{Offset}} &
   \multicolumn{1}{l}{\textbf{$V_P$}}  &
   \multicolumn{1}{l}{\textbf{fwhm}} &
   \multicolumn{1}{l}{\textbf{Flux \footnotemark[1]}} &  
   \multicolumn{1}{l}{\textbf{Offset}} &
   \multicolumn{1}{l}{\textbf{$V_P$}} &
   \multicolumn{1}{l}{\textbf{fwhm}} &
   \multicolumn{1}{l}{\textbf{Flux}} & 
   \multicolumn{1}{l}{\textbf{Flux}}  \\[0.8ex]
   \multicolumn{1}{l}{\textbf{}} &
   \multicolumn{1}{l}{\textbf{}} &
   \multicolumn{1}{l}{\textbf{arcsec}} &
   \multicolumn{1}{l}{\textbf{\,\kms}} &
   \multicolumn{1}{l}{\textbf{\,\kms}} &
   \multicolumn{1}{l}{\textbf{cgs}} &  
   \multicolumn{1}{l}{\textbf{arcsec}} &
   \multicolumn{1}{l}{\textbf{\,\kms}} &
   \multicolumn{1}{l}{\textbf{\,\kms }} &
   \multicolumn{1}{l}{\textbf{cgs}} & 
   \multicolumn{1}{l}{\textbf{cgs}}\\[0.5ex] \hline
   \\[-1.8ex]
\endhead 
\multicolumn{1}{l}{\textbf{}} &
\multicolumn{1}{l}{\textbf{}} &
\multicolumn{1}{c}{\textbf{Period 1}} \\[0.5ex] \hline
\\[-1.8ex] 
(1) 2002-01-28  &  $S_{-2b}$ 		  & 0.185  & -241  & 84.0 &  1.6e-13 &   --    &  --  & -- & -- 	   & 4.0e-13 \\
            &  $S_{0b}$  		  & 0.160  & -180  & 85.1 &  2.2e-13 &   --    &  --  & -- & -- 	   & 4.2e-12 \\
            &  $S_{+2b}$		  & --      &  --   & --    &   --     &   --    &  --  & -- & -- 	   & 1.7e-13\\
\hline
(2) 2002-12-29  &  $S_{-2t}$		  & 0.312  & -220  & 74.4 &  4.6e-14 &   --    &  --  & -- & -- 	   & 1.4e-13\\
            &  $S_{-1t}$		  & 0.236  & -211  & 77.8 &  2.1e-13 &   --    &  --  & -- & -- 	   & 7.7e-13 \\
            &  $S_{0t}$ 		  & 0.236  & -180  & 58.4 &  5.1e-14 &   --    &  --  & -- & -- 	   & 5.4e-12\\
            &  $S_{+1t}$ 		  &  --     &  --   & --    &   --     &   --    &  --  & -- & -- 	   & 5.5e-13\\
            &  $S_{+2t}$		  &  --     &  --   & --    &   --     &   --    &  --  & -- & -- 	   & -- \\
\hline
(3) 2004-01-12  &  $S_{-1t}$		  & 0.313  & -200  & 82.6 &  3.4e-13 &   --    &   -- & --    & --  	   & 8.8e-13\\ 
	    &  $S_{0t}$ 		  & 0.365  &  -176 & 94.9 & 5.5e-14  & 0.056  & -139 & 99.70 & 7.9e-13 & 5.6e-12\\ 
	    &  $S_{+1t}$		  &  --     &  --   & --    &  --      &   --    &   -- & --    & -- 	   & 4.2e-13\\
	    &  $S_{+2t}$		  &  --     &  --   & --    &   --     &   --    &  --  & --    & -- 	   & -- \\
\hline \\[-2ex]
\multicolumn{1}{l}{\textbf{}} &
\multicolumn{1}{l}{\textbf{}} &
\multicolumn{1}{c}{\textbf{Period 2}} \\[0.5ex] \hline
\\[-1.8ex]
(4) 2011-07-07  &  $S_{+2b}$	          & 0.236  & -180  & 69.5 &  2.5e-14 & --  & -- & -- & -- & 6.0e-14\\
	    &  $S_{0b}$   &  --\footnotemark[2]     &  --\footnotemark[2]   & --\footnotemark[2]    &    --    & 0.033  & -160 & 70.86 & 2.2e-13 & 1.0e-12 \\
	    &  $S_{-2b}$ 		  &  --     &  --   & --    &    --    &   --    &  --  & --    & -- 	   & 1.4e-14 \\
\hline
(5) 2012-07-14  &  $S_{+2b}$		  &  0.312 &  -180 & 43.9 & 1.4e-14  &  0.008 & -200 & 122.24 & 2.4e-14 &   -- \\
	    &  $S_{0b}$ &   --\footnotemark[2]    &  --\footnotemark[2]   & --\footnotemark[2]    &    --    &  0.033 & -200 & 99.48  & 8.2e-13 &   7.6e-13\\
	    &  $S_{-2b}$ 		  &  --     &  --   & --    &   --     &  0.058 & -180 & 138.35 & 2.2e-14 &   4.8e-14 \\
\hline
(6) 2013-07-17  &  $S_{+2b}$ 		  & 0.363  & -160  & 63.1 &  1.2e-14 &   --    & --   & --     &   --    &  -- \\
	    &  $S_{0b}$ 		  & 0.36   & -157  & 46.4 &  1.3e-14 &  0.071 & -180 & 74.88  & 5.3e-13 & 1.5e-12\\
	    &  $S_{-2b}$  		  &  --    &  --   & --    &   --     &   --    &  --  & --     &   --    & --
\label{obssii}
\end{longtable}
\end{turnpage}

\footnotetext[1]{Intensity units (cgs) are erg\,s$^{-1}$\,cm$^{-2}$\,{\AA}$^{-1}$\,arcsec$^{-2}$}
\footnotetext[2]{We can see emission but it is not separated clearly from the central blob, so we cannot measure a spatial location.}



\clearpage
\scriptsize
\begin{turnpage}
\begin{longtable}{p{0.8in}p{0.45in}p{0.45in}p{0.45in}p{1.0in}}
\caption[]{Observational Properties Ib -- Distant Blobs ([SII] $\lambda$4069.75)}\\
\hline \\[-2ex]
   \multicolumn{1}{l}{\textbf{Epoch}} &
   \multicolumn{1}{l}{\textbf{Slit}} &
   \multicolumn{1}{l}{\textbf{Offset}} &
   \multicolumn{1}{l}{\textbf{$V_P$}} &
   \multicolumn{1}{l}{\textbf{Flux}} \\[0.5ex] 
   \multicolumn{1}{l}{\textbf{(\#) Date}} &
   \multicolumn{1}{l}{\textbf{Name}} &
   \multicolumn{1}{l}{\textbf{arcsec}} &
   \multicolumn{1}{l}{\textbf{\,\kms}} &
   \multicolumn{1}{l}{\textbf{cgs}} \\[0.5ex] 
   \\[-1.8ex]
\endfirsthead
\multicolumn{3}{c}{{\tablename} \thetable{} -- Continued} \\[0.5ex]
\hline \\[-2ex]
   \multicolumn{1}{l}{\textbf{Epoch}} &
   \multicolumn{1}{l}{\textbf{Slit}} &
   \multicolumn{1}{l}{\textbf{Offset}} &
   \multicolumn{1}{l}{\textbf{$V_P$}} &
   \multicolumn{1}{l}{\textbf{Flux}} \\[0.5ex] 
   \multicolumn{1}{l}{\textbf{(\#) Date}} &
   \multicolumn{1}{l}{\textbf{Name}} &
   \multicolumn{1}{l}{\textbf{arcsec}} &
   \multicolumn{1}{l}{\textbf{\,\kms}} &
   \multicolumn{1}{l}{\textbf{cgs}} \\[0.5ex] 
   \\[-1.8ex]
\endhead
\hline \\[-2ex]
\multicolumn{1}{l}{\textbf{}} &
\multicolumn{1}{l}{\textbf{}} &
\\[-1.8ex]
(1) 2002-01-28  &  $S_{+2b}$    &  1     &  -161   &  8.7e-15   \\
(4) 2011-07-07\footnotemark[1]  &  $S_{-2b}$  & 0.75   &  -172   &  8.1e-15   \\
(5) 2012-07-14\footnotemark[1]  &  $S_{-2b}$  & 0.8   &  -172   &  8.1e-15   \\ 
\label{obsdist}
\end{longtable}
\end{turnpage}
\footnotetext[1]{We have averaged these datasets together, due to the relatively low S/N, 
for determining the radial velocity and emission intensity.}

\clearpage

\scriptsize
\begin{turnpage}
\begin{longtable}{p{0.8in}p{0.35in}p{0.35in}p{0.35in}p{0.9in}}
\caption[]{Observational Properties II -- Detached \& On-Source Blobs ([FeII] $\lambda$4245.16)}\\
\hline \\[-2ex]
   \multicolumn{1}{l}{\textbf{Epoch}} &
   \multicolumn{1}{l}{\textbf{Slit}} &
   \multicolumn{1}{l}{\textbf{Offset}} &
   \multicolumn{1}{l}{\textbf{$V_P$}} &
   \multicolumn{1}{l}{\textbf{Flux}} \\[0.5ex] 
   \multicolumn{1}{l}{\textbf{(\#) Date}} &
   \multicolumn{1}{l}{\textbf{Name}} &
   \multicolumn{1}{l}{\textbf{arcsec}} &
   \multicolumn{1}{l}{\textbf{\,\kms}} &
   \multicolumn{1}{l}{\textbf{cgs}} \\[0.5ex] 
   \\[-1.8ex]
\endfirsthead
\multicolumn{3}{c}{{\tablename} \thetable{} -- Continued} \\[0.5ex]
\hline \\[-2ex]
   \multicolumn{1}{l}{\textbf{Epoch}} &
   \multicolumn{1}{l}{\textbf{Slit}} &
   \multicolumn{1}{l}{\textbf{Offset}} &
   \multicolumn{1}{l}{\textbf{$V_P$}} &
   \multicolumn{1}{l}{\textbf{Flux}} \\[0.5ex] 
   \multicolumn{1}{l}{\textbf{(\#) Date}} &
   \multicolumn{1}{l}{\textbf{Name}} &
   \multicolumn{1}{l}{\textbf{arcsec}} &
   \multicolumn{1}{l}{\textbf{\,\kms}} &
   \multicolumn{1}{l}{\textbf{cgs}} \\[0.5ex] 
   \\[-1.8ex]
\endhead
\hline \\[-2ex]
\multicolumn{1}{l}{\textbf{}} &
\multicolumn{1}{l}{\textbf{}} &
\multicolumn{1}{c}{\textbf{Period 1: Detached Blob}} \\[0.5ex] \hline
\\[-1.8ex]    
(1) 2002-01-28  &  $S_{-2b}$  & 0.211  & -236& 2.1e-14 \\
(2) 2002-12-29  &  $S_{-1t}$  & 0.262  & -219 & 3.9e-14   \\
(3) 2004-01-12  &  $S_{-1t}$  & 0.262  &  -199  &  2.5e-14 \\
\hline \\[-2ex]
\multicolumn{1}{l}{\textbf{}} &
\multicolumn{1}{l}{\textbf{}} &
\multicolumn{1}{c}{\textbf{Period 2: On-Source Blob}} \\[0.5ex] \hline
\\[-1.8ex]
(4) 2011-07-07  &  $S_{0b}$   & ...\footnotemark[1]  & -169 & 5.2e-14   \\
(5) 2012-07-14  &  $S_{0b}$   & ...\footnotemark[1] &  -183  & 3.2e-13  \\
(6) 2013-07-17  &  $S_{0b}$   & ...\footnotemark[1]  & -161  & 1.2e-13 \\
\label{obsfe1}
\end{longtable}
\end{turnpage}
\footnotetext[1]{Blob offset cannot be measured reliably due to imperfect subtraction of the relatively strong underlying continuum at the Fe\,II line wavelength.}

\clearpage
\scriptsize
\begin{longtable}{p{0.8in}p{0.45in}p{0.45in}p{0.45in}p{1.0in}}
\caption[]{Observational Properties III - Detached \& On-Source Blobs ([FeII] $\lambda$4288.6)}\\
\hline \\[-2ex]
   \multicolumn{1}{l}{\textbf{Epoch}} &
   \multicolumn{1}{l}{\textbf{Slit}} &
   \multicolumn{1}{l}{\textbf{Offset}} &
   \multicolumn{1}{l}{\textbf{$V_P$}} &
   \multicolumn{1}{l}{\textbf{Flux}} \\[0.5ex] 
   \multicolumn{1}{l}{\textbf{(\#) Date}} &
   \multicolumn{1}{l}{\textbf{Name}} &
   \multicolumn{1}{l}{\textbf{arcsec}} &
   \multicolumn{1}{l}{\textbf{\,\kms}} &
   \multicolumn{1}{l}{\textbf{cgs}} \\[0.5ex] 
   \\[-1.8ex]
\endfirsthead
\multicolumn{3}{c}{{\tablename} \thetable{} -- Continued} \\[0.5ex]
\hline \\[-2ex]
   \multicolumn{1}{l}{\textbf{Epoch}} &
   \multicolumn{1}{l}{\textbf{Slit}} &
   \multicolumn{1}{l}{\textbf{Offset}} &
   \multicolumn{1}{l}{\textbf{$V_P$}} &
   \multicolumn{1}{l}{\textbf{Flux}} \\[0.5ex] 
   \multicolumn{1}{l}{\textbf{(\#) Date}} &
   \multicolumn{1}{l}{\textbf{Name}} &
   \multicolumn{1}{l}{\textbf{arcsec}} &
   \multicolumn{1}{l}{\textbf{\,\kms}} &
   \multicolumn{1}{l}{\textbf{cgs}} \\[0.5ex] 
   \\[-1.8ex]
\endhead
\hline \\[-2ex]
\multicolumn{1}{l}{\textbf{}} &
\multicolumn{1}{l}{\textbf{}} &
\multicolumn{1}{c}{\textbf{Period 1: Detached Blob}} \\[0.5ex] \hline 
\\[-1.8ex]
(1) 2002-01-28  &  $S_{-2b}$  &  0.185   &  -246     & 1.8e-14  \\
(2) 2002-12-29  &  $S_{-1t}$  &  0.236    & -220     & 4.5e-14  \\
(3) 2004-01-12  &  $S_{-1t}$  &  0.338    &  -196    & 4.0e-14   \\
\hline \\[-2ex]
\multicolumn{1}{l}{\textbf{}} &
\multicolumn{1}{l}{\textbf{}} &
\multicolumn{1}{c}{\textbf{Period 2: On-Source Blob}} \\[0.5ex] \hline
\\[-1.8ex]
(4) 2011-07-07  &  $S_{0b}$   & ...\footnotemark[1]  &  -166    &  2.8e-14   \\
(5) 2012-07-14  &  $S_{0b}$   & ...\footnotemark[1]  &  -195   &  1.9e-13   \\
(6) 2013-07-17  &  $S_{0b}$   & ...\footnotemark[1]  &  -163   &  7.8e-14  \\    
\label{obsfe2}
\end{longtable}
\footnotetext[1]{Blob offset cannot be measured reliably due to imperfect subtraction of the relatively strong underlying continuum at the Fe\,II line wavelength.}


\clearpage
\begin{table}[!t]\label{tabflipflop}
\caption{Observational Characteristics Associated with the Flip-Flop Phenomenon}
\label{obs-parm}
\begin{tabular}{lllllll}
\hline
Bullet & Ejection  & Light-Curve            & Ejection  & Obs.   & Blob & Radial Vel. \\
No.  & Epoch     & Phase\tablenotemark{a} & Direction & Epoch  & Type & ($\kms$)  \\
\hline
0 & (1986-1988)\tablenotemark{b} &   $0.04$    & ENE     & 2002.1      & distant    & $-161$  \\
  &                              &             &         & 1986\,Dec--1990\,Mar\tablenotemark{c} &  & $-166$  \\
1 & (1993.5-1995.5)\tablenotemark{b}    &   $0.51$   & ESE     & 2002.1,2003,2004   & detached  & $-241,-211,-200$ \\
  &              &             & ESE     & 2011.6-2012.6 & distant  & $-172$ \\
2 & 2002.1-2004    &  $0.87-0.99$ & (ENE)\tablenotemark{b} & 2004        & on-source  & $-140$ \\
  &              &              & ENE     & 2011.6,2012.6,2013.6 & detached & $-180,-180,-160$ \\
3 & 2011.6       & $1.43-1.55$ & (ESE)\tablenotemark{b}  & 2011.6,2012.6,2013.6 & on-source  & $-160,-200,-180$ \\
\hline
\tablenotetext{1}{Phase of the 6160\,d periodic variation in the light-curve of V\,Hya (0 implies maximum light)}
\tablenotetext{2}{Values in parenthesis are inferred}
\tablenotetext{3}{Ground-based observations obtained by Lloyd Evans (1991)}
\end{tabular}
\end{table}

\clearpage

\clearpage

\clearpage

\end{document}